\documentclass[aip,jcp,citeautoscript,amsmath,amssymb,preprint]{revtex4-2}
\usepackage{xcolor}
\usepackage{hyperref}
\usepackage[version=4]{mhchem}
\usepackage{blindtext}
\usepackage{graphicx}
\usepackage[skip=0.333\baselineskip]{caption}
\usepackage{subcaption}
\usepackage{adjustbox}
\usepackage{siunitx}
\usepackage{comment}

\colorlet{review}{red}
\makeatletter
\def\@email#1#2{%
 \endgroup
 \patchcmd{\titleblock@produce}
  {\frontmatter@RRAPformat}
  {\frontmatter@RRAPformat{\produce@RRAP{*#1\href{mailto:#2}{#2}}}\frontmatter@RRAPformat}
  {}{}
}%
\makeatother
\begin{document}

\title[]{Enhancing the Accuracy of XPS Calculations: Exploring Hybrid Basis Set Schemes for CVS-EOMIP-CCSD Calculations}

\author{Alexis A. A. Delgado}
 \affiliation{Department of Chemistry, Southern Methodist University, Dallas, TX 75275, USA}
\author{Devin A. Matthews*}
 \affiliation{Department of Chemistry, Southern Methodist University, Dallas, TX 75275, USA}
 \email{damatthews@smu.edu}

\begin{abstract}
Reliable computational methodologies and basis sets for modeling x-ray spectra are essential for extracting and interpreting electronic and structural information from experimental x-ray spectra. In particular, the trade-off between numerical accuracy and computational cost due to the size of the basis set is a major challenge, since molecular orbitals undergo extreme relaxation in the core-hole state. To gain clarity on changes in electronic structure induced by the formation of a core-hole, the use of sufficiently flexible basis for expanding the orbitals, particularly for the core region, has been shown to be essential. This work focuses on the refinement of core-hole ionized state calculations using the equation-of-motion coupled cluster (EOM-CC) family of methods through an extensive analysis on the effectiveness of ``hybrid" and mixed basis sets. In this investigation, we utilize the CVS-EOMIP-CCSD method in combination and construct hybrid basis sets piece-wise from readily available Dunning's correlation consistent basis sets in order to calculate x-ray ionization energies for a set of small gas phase molecules. Our results provide insights into the impact of basis sets on CVS-EOMIP-CCSD calculations of K-edge ionization energies of first-row p-block elements. These insights enable us to understand more about the basis set dependence of the core IEs computed and allow us to establish a protocol for deriving reliable and cost-effective theoretical estimates for computing IEs of small molecules containing such elements.
\end{abstract}

\maketitle

\section{Introduction}

X-ray spectroscopy techniques take advantage of the highly localized and element-specific nature of core orbitals to provide in-depth knowledge of molecular and electronic structure. The creation of a core hole, due to the ejection of core electrons by x-rays into the vacuum (x-ray photoelectron spectroscopy, XPS) or high-lying virtual orbitals (x-ray absorption spectroscopy, XAS, specifically near-edge x-ray absorption fine struct, NEXAFS, or x-ray absorption near-edge spectroscopy, XANES) severely disrupts the electron configuration of the system. The creation of a K-edge core hole increases the effective nuclear charge and gives rise to extreme orbital relaxation effects. To devise accurate theoretical models for x-ray spectroscopy, such orbital relaxation effects must be adequately addressed at the self-consistent field and/or electron correlation levels as well as in the flexibility of the orbital basis set.

The theoretical frameworks underlying the computation of XPS and XAS spectra can be divided into groups which either (a) access the ground and core-ionized/excited states directly in separate calculations or (b) use response/propagator theory from the ground state to access core ionized/excited states. 
Some examples of methods of type (a) include ``$\Delta$" Hartree--Fock ($\Delta$HF)\cite{Bagus1965, NavesdeBrito1991, Schmitt1992, Besley2009} and Kohn--Sham density functional theory ($\Delta$KS, collectively $\Delta$SCF),\cite{Besley2009, Triguero1999} restricted active-space self-consistent-field theory (RASSCF),\cite{NavesdeBrito1991,Jensen1987,Aagren1994,Bagus2016} orthogonality-constrained density functional theory (OC-DFT),\cite{Verma2015} static exchange (STEX),\cite{Aagren1994, Ekstroem2006} and non-orthogonal configuration interaction singles (NOCIS).\cite{Oosterbaan2018} Examples of methods of type (b) include configuration interaction singles (CIS, or the related time-dependent Hartree--Fock method, TD-HF),\cite{Maganas2014, Toffoli2016, Ehlert2016} CIS with perturbative inclusion of double excitations (CIS(D)),\cite{Asmuruf2008} time-dependent (linear response) DFT (TD-DFT) theory,\cite{DeBeerGeorge2008, Besley2009} the Bethe-Salpeter equation approach,\cite{Vinson2011} Green’s function methods,\cite{Barth1985, Barth1981, Trofimov2000, Wenzel2015} equation-of-motion coupled-cluster (EOM-CC) methods\cite{Southworth2015, Peng2015} or the closely related coupled-cluster linear response (CC-LR) methods,\cite{Coriani2012,Coriani2015} and multi-reference coupled-cluster methods.\cite{Datta2009, Brabec2012, Sen2013, Dutta2014} In particular, equation-of-motion coupled-cluster (EOM-CC) theory allows for a balanced and multi-reference treatment of excited or ionized states in the same Fock sector and effectively treats the electron correlation effects beyond the Hartree-Fock approximation.\cite{Emrich1981, Stanton1993, Krylov2008, Bartlett2011, Sneskov2011} Because of their properties, EOM-CC methods, with the incorporation of the core-valence separation (CVS) scheme,\cite{Coriani2015, Cederbaum1980, Coriani2016, Vidal2019, Tenorio2019, Liu2019} are popular amongst the methods used for modeling spectroscopic phenomena of core-level states. Various benchmarks have shown that CVS-EOM-CC methods accurately calculate core-ionized/excited states.\cite{Coriani2015, Coriani2016, Vidal2019, Tenorio2019, Liu2019, Myhre2016, Frati2019}


In addition to the type of computational method used to compute core-level spectra, the basis set used to construct the molecular orbitals of the ground and core-ionized/excited states also influences the accuracy of the results. Because a majority of commonly used basis sets were primarily developed for calculating ground or valence excited state energies, properties, geometries, analytical frequencies, etc. it is unsurprising that accurate prediction of other molecular properties (i.e. core-ionization/core-excitation energies, J coupling constants, isotropic hyperfine coupling constants, isotropic shielding constants, etc.) may require customized basis sets.\cite{Jensen2021,Ireland2023} Since core electrons, as well as the core hole, are tightly bound to the nucleus their molecular orbitals are highly localized and, particularly in the case of the core hole, not well represented by typical contracted basis functions in common basis sets. Additionally, the contraction of the occupied valence and, to a lesser extent, the virtual orbitals upon creation of a core hole requires greater radial flexibility than most standard basis sets provide. Finally, in linear response-type methods, the basis set must simultaneously be able to accurately describe the ``normal" ground state orbitals as well as the relaxed orbitals in the core-hole state. A number of studies have explored these issues, as well as several suggested strategies for mitigating them by selecting appropriate basis sets or even creating customized basis sets.\cite{Besley2009, Watson2013, Mijovilovich2009, Carbone2019,Shirai2004,Takahata2003,Wenzel2014,Wenzel2015,Kovac2014,Fransson2016,Tolbatov2017,Fouda2017,Hanson-Heine2018,Ambroise2018,Hait2020}

A little over two decades ago, Cavigliasso et al. carried out a study on the basis set dependence of $\Delta$KS calculations for K-edge (1s) core-electron ionization energies (IEs, or equivalently core-electron binding energies) of 17 small molecules comprising first-row p-block elements.\cite{Cavigliasso1999} They found the standard cc-pCVTZ basis to be as effective as the cc-pV5Z basis as the cc-pCVTZ basis set provided greater completeness in the space spanned by the basis set due to the inclusion of tight core functions. An investigation by Besley et al. explored the use of uncontracted basis sets with $\Delta$SCF calculations and found that IEs computed with the uncontracted 6-311G** basis provided results with similar quality obtained with the contracted cc-pCVQZ basis.\cite{Besley2009} In the work of Tolbatov and co-workers, basis set performance on C, N, and O 1s IEs of amino acids, computed with commonly used density functionals, were investigated and the results suggested that the PBE0/6-311+G* method performed best.\cite{Tolbatov2017} Around the same time, Fouda et al. evaluated the basis set convergence of 1s core IEs for 34 small molecules computed with $\Delta$SCF and found that the cc-pCVTZ basis set yielded accurate results.\cite{Fouda2017} Govind and coworkers showed that the cc-pV$n$Z and def2-$n$ZVP basis sets are not well suited for computing IEs at the GW level of theory due to contraction errors. Their calculations revealed that uncontracted versions of the the cc-pV$n$Z and def2-$n$ZVP basis sets improved convergence and that faster convergence could be achieved using uncontracted pcJ-$n$, pcSseg-$n$, and ccX-$n$Z basis sets.\cite{Mejia-Rodriguez2022}

A different tack was taken by Hanson-Heine et al., where they calculated $\Delta$SCF 1s core IEs by supplementing small standard basis sets, 6-31G* and cc-pVDZ, for an element of nuclear charge $Z$ with basis functions from the element with charge $Z+1$.\cite{Hanson-Heine2018} Their work revealed that IEs obtained at the PBE/$Z+1$(6-31G*) level of theory fell within 0.5 eV of their reference IEs obtained at the DFT/cc-pCVQZ level of theory, and that using the $Z+1$ basis set only for ``active" atom (where the 1s core orbital is localized) delivered sufficient results with fewer basis functions.\cite{Hanson-Heine2018}  Ambroise et al. \cite{Ambroise2018} investigated basis set requirements for computing XPS for N, F, and C K-edges of small molecules using $\Delta$SCF. They defined a series of pcX-$n$ basis sets based on the $Z+1$ construct previously mentioned, but utilizing ``$Z+1/2$" functions with interpolated exponents (from $Z$ to $Z+1$), and acquired near-optimum basis function exponents that provided a balanced description for ground and core-ionized states, greatly lowered basis set errors, and reduced computational cost. They also observed a monotonic and smooth convergence of the IEs with standard basis sets and found that the aug-cc-pCVTZ basis set provided core IEs within 0.1 eV of the results obtained with the aug-cc-pCV5Z basis set.\cite{Ambroise2018} The same authors found that the pcX-$n$ basis sets were not effective for CVS-EOM-CC methods, and so they further explored a set of ccX-$n$Z basis sets which were created by combining custom, lightly contracted $sp$ sets with core (valence) polarization functions derived from standard basis sets with cardinal number $n+1$ ($n$). These basis sets were found to provide smaller basis set errors in comparison to commonly used basis sets (of equivalent size) where the basis set error for the smallest ccX-$n$Z basis set (i.e. ccX-DZ) was around 0.1 eV.\cite{Ambroise2021}

Sarangi and co-workers computed an extensive benchmark of core IEs for C, N, and O K-edges of small molecules (primarily diatomic, with two polyatomic), using the fc-CVS-EOMIP-CCSD method which employs the frozen-core approximation in the ground state calculation.\cite{Sarangi2020} Contracted, core-uncontracted and fully-uncontracted Pople and Dunning's standard basis sets (with and without tight core functions), as well as mixed basis set schemes (which apply different basis sets to different atoms in the molecule) were employed up to quintuple-$\zeta$ quality. Their results suggested that using an uncontracted basis for the atom to be ionized and smaller standard basis sets for the remainder of the molecule did not significantly impact the core IEs. Also, they observed that the basis sets used for the H atoms had little to no impact on computed core IEs. 
Though they found that the uncontracted 6-311G+(3df) basis set greatly reduced basis set error and performed similarly to the standard aug-cc-pV5Z basis,\cite{Sarangi2020} Ambroise et al. noted that their ccX-DZ basis performs just as well the uncontracted 6-311G+(3df) basis but is much smaller in size, thereby diminishing computational cost.\cite{Ambroise2021} 

These previous results all illustrate a need for more flexibility in the basis set either by using very large standard basis sets (not preferred for efficient calculations), including tight core functions (almost always advisable), diffuse augmented basis sets (not obviously appropriate for ionization processes), core- or fully-uncontracted basis sets (potentially effective, particularly if applied only to the active atom), or using customized contracted basis sets (most preferred in terms of basis set size). In this paper, we explore a technique for combining basis sets following several of these guidelines, which however eschews customized exponents or contraction schemes. Instead, the present approach utilizes only contracted, off-the-shelf basis functions from standard Dunning's basis sets to construct ``hybrid" basis sets for obtaining reliable IEs at the CVS-EOMIP-CCSD level. The overall goals of this research are:
\begin{itemize}
    \item to assess the accuracy and efficiency of off-the-shelf hybrid basis sets in comparison to previously-published approaches;
    \item to more thoroughly investigate the use of mixed basis sets as a means of lowering computational cost without decreasing accuracy;
    \item to understand the role of augmented (diffuse) functions; and
    \item to explore the effect of freezing non-active core orbitals (while simultaneously removing the associated tight core functions);
\end{itemize}
in all cases in the context of core IEs computed at the CVS-EOMIP-CCSD level. By using portions of existing and easily-obtainable basis sets in place of specially-optimized basis sets for core-level properties, we hope to create a basis set protocol which is immediately extensible beyond the typical first- and second-row elements.

\section{Computational Details}

Our test set includes 1s principal core ionization energies of all heavy atoms within the following molecules: \ce{C2H4}, \ce{H2O}, \ce{HCN}, \ce{NH3}, \ce{CO}, \ce{CH2}, \ce{CH4}, \ce{H2CNH}, \ce{H2NF}, \ce{H2CO}, \ce{H3CF}, \ce{H3COH}, \ce{HF}, \ce{HNO}, and \ce{HOF}. This test set consists of 24 vertical ionization energies (the average of the tunneling levels for \ce{C2H4} is used). The core-valence separated equation-of-motion coupled cluster CVS-EOMIP-CCSD method,\cite{Liu2019} with and without frozen core applied (as indicated in the text), was combined with standard, off-the-shelf Dunning's correlation consistent basis sets cc-pV$n$Z (cardinal number $n$ = D, T, Q, 5, 6, or numerically as $n$ = 2--6), optionally with diffuse (aug) and/or tight core (C) functions.\cite{dunning1989gaussian, kendall1992electron, dunning1993gaussian, woon1995gaussian, wilson1996gaussian, van1999benchmark, peterson2002accurate} In particular, tight core functions are used for all heavy atoms whose core 1s electrons are not frozen and for the uncontracted basis sets used for comparison in this work. In addition to completely uncontracted basis sets (uncontracted cc-pCV$n$Z, denoted $\text{u}n$), we also employ core-uncontracted cc-pCV$n$Z basis sets, denoted $\text{uC}n$. Because the $1s$ and $2s$ orbitals share primitives in the Dunning basis sets, this amounts to uncontracting all $s$ functions. In addition, the ccX-$n$Z ($n$ = D, T, Q, 5) basis sets, taken from the work of Ambroise et al.,\cite{Ambroise2021} were also used for comparison. All calculations were preformed using the CFOUR program package.\cite{Matthews2020} Reference IEs for the test set were obtained at the CVS-EOMIP-CCSD/aug-cc-pCV6Z level of theory with the exception of molecules \ce{H3COH} and \ce{C2H4} whose reference IEs were computed at the CVS-EOMIP-CCSD/aug-cc-pCV5Z level of theory. For molecules where sextuple-$\zeta$ results were obtained, the difference with respect to quintuple-$\zeta$ is in all cases less than 10 meV.

\begin{table}\renewcommand*{\arraystretch}{0.65}
\begin{tabular}{c|cccc|cc}
 & \multicolumn{4}{c}{CNOF} & \multicolumn{2}{|c}{H} \\
Basis Set & $n$Z & C$n$Z & a$n$Z & aC$n$Z & $n$Z & a$n$Z \\
  \hline
2/2 &  14 &  18 &  23 &  27 &  5 &   9 \\
3/2 &  18 &  26 &  27 &  35 &  5 &   9 \\
4/2 &  22 &  34 &  31 &  43 &  5 &   9 \\
5/2 &  26 &  42 &  35 &  51 &  5 &   9 \\
6/2 &  30 &  50 &  39 &  59 &  5 &   9 \\
3/3 &  30 &  43 &  46 &  59 & 14 &  23 \\
4/3 &  34 &  51 &  50 &  67 & 14 &  23 \\
5/3 &  38 &  59 &  54 &  75 & 14 &  23 \\
6/3 &  42 &  67 &  58 &  83 & 14 &  23 \\
4/4 &  55 &  84 &  80 & 109 & 30 &  46 \\
5/4 &  59 &  92 &  84 & 117 & 30 &  46 \\
6/4 &  63 & 100 &  88 & 125 & 30 &  46 \\
5/5 &  91 & 145 & 127 & 181 & 55 &  80 \\
6/5 &  95 & 153 & 131 & 189 & 55 &  80 \\
6/6 & 140 & 230 & 189 & 279 & 91 & 127 \\
\hline
uC2 & --- &  24 &  --- &  --- & 5  & --- \\ 
uC3 & --- &  49 &  --- &  --- & 14 & --- \\ 
uC4 & --- &  91 &  --- & --- & 30 & --- \\ 
uC5 & --- & 153 & --- & --- & 55 & --- \\ 
\hline 
u2 & ---  &  30 & ---  & ---  & 5 & --- \\
u3 & ---  &  55 & ---  & ---  & 14 & --- \\ 
u4 & --- &  97 &  --- & --- & 30 & ---\\ 
u5 & --- & 162 & --- & --- & 55 & --- \\ 
\hline
ccX-DZ & --- & 38  & --- & --- &  5 & --- \\
ccX-TZ & ---  & 70 & --- & --- & 14 & --- \\ 
ccX-QZ & ---  & 120 & --- & --- & 30 & --- \\ 
ccX-5Z & --- & 192 & --- & --- & 55 & --- \\ 
\end{tabular}
\caption{Number of basis functions on heavy atoms (CNOF) or hydrogen for hybrid ($n$/$m$), core-uncontracted (uC$n$), fully uncontracted (u$n$), and ccX-$n$Z basis sets. The short-hand notation $n$Z~=~cc-pV$n$Z, a$n$Z~=~aug-cc-pV$n$Z, C$n$Z~=~cc-pCV$n$Z, aC$n$Z~=~aug-cc-pCV$n$Z is used. Basis functions for hydrogen are kept contracted in all cases. \label{tab:hybrid}}
\end{table}

Our hybrid basis sets were constructed by combining the (aug-)cc-p(C)V$n$Z $s$ and $p$ functions with the valence polarization ($d$ and higher angular momentum) functions from the matching (aug-)cc-p(C)V$m$Z basis sets such that $2 \le m < n \le 6$. For hydrogen, the (aug-)cc-pV$m$Z basis set is used. Such hybrid combinations are denoted briefly as $n$/$m$, so that for example 4/2 denotes a $5s4p1d$ basis set which is a mixture of cc-pVQZ $sp$ functions with cc-pVDZ $d$ functions (and diffuse/core functions as indicated separately). The standard (non-hybrid) basis sets may be indicated as either $n$/$m$ or simply $n$. In addition to hybrid basis sets which are applied evenly across the molecule, we tested several types of mixed basis sets which utilize different sets on different classes of atoms. A convenient short-hand notation for such combinations is [$n$/$m$, $n^\prime$/$m^\prime$, $m^\ast$], where basis set $n$/$m$ is applied to the active atom (where the core hole is localized), basis set $n^\prime$/$m^\prime$ is applied to all other heavy atoms, and basis set $m^\ast$ (always a standard basis set (aug-)cc-pV$m^\ast$Z) is applied to hydrogen. In general these five cardinal numbers can be any numbers, but we focus specifically on the cases $2 \le m^\prime = n^\prime = m^\ast < m = n \le 6$ (standard mixed) and $2 \le m^\prime = n^\prime = m^\ast = m < n \le 6$ (hybrid mixed). When necessary for disambiguation, we will also refer to aug-[$n$/$m$, $n^\prime$/$m^\prime$, $m^\ast$] and [aug-$n$/$m$, $n^\prime$/$m^\prime$, $m^\ast$] basis sets, where in the former diffuse functions are added on all atoms and in the latter diffuse functions are added only to the active atom. Otherwise, the presence of diffuse functions will be inferred from context. The hybrid (and for $n=m$, standard) basis sets used in this work are summarized in Table~\ref{tab:hybrid} along with the number of basis functions per atom.

\section{Results and Discussion}

The data set for our hybrid basis set type schemes consists of all 24 vertical IEs across our entire test set whereas the data set for the standard mixed and hybrid mixed basis set type schemes consists of 18 vertical IEs for the molecules of our test set that contain two (non-equivalent) heavy atoms (\ce{HCN}, \ce{CO}, \ce{H2CNH}, \ce{H2NF}, \ce{H2CO}, \ce{H3CF}, \ce{H3COH}, \ce{HNO}, and \ce{HOF}).

Relative IE errors for our basis sets, for every K-edge computed, are calculated by subtracting the IE computed with our basis set from our reference IE, IE(B) - IE(REF). For each specific basis set combination, we present both raw data for each IE, as well as statistical summary based on the 95\% confidence interval of the error (assuming a Gaussian distribution of errors) and the average number of basis functions (b.f.s) per atom. The latter number varies across the test set depending on the saturation of each compound.

\begin{figure}
\includegraphics[width=\textwidth]{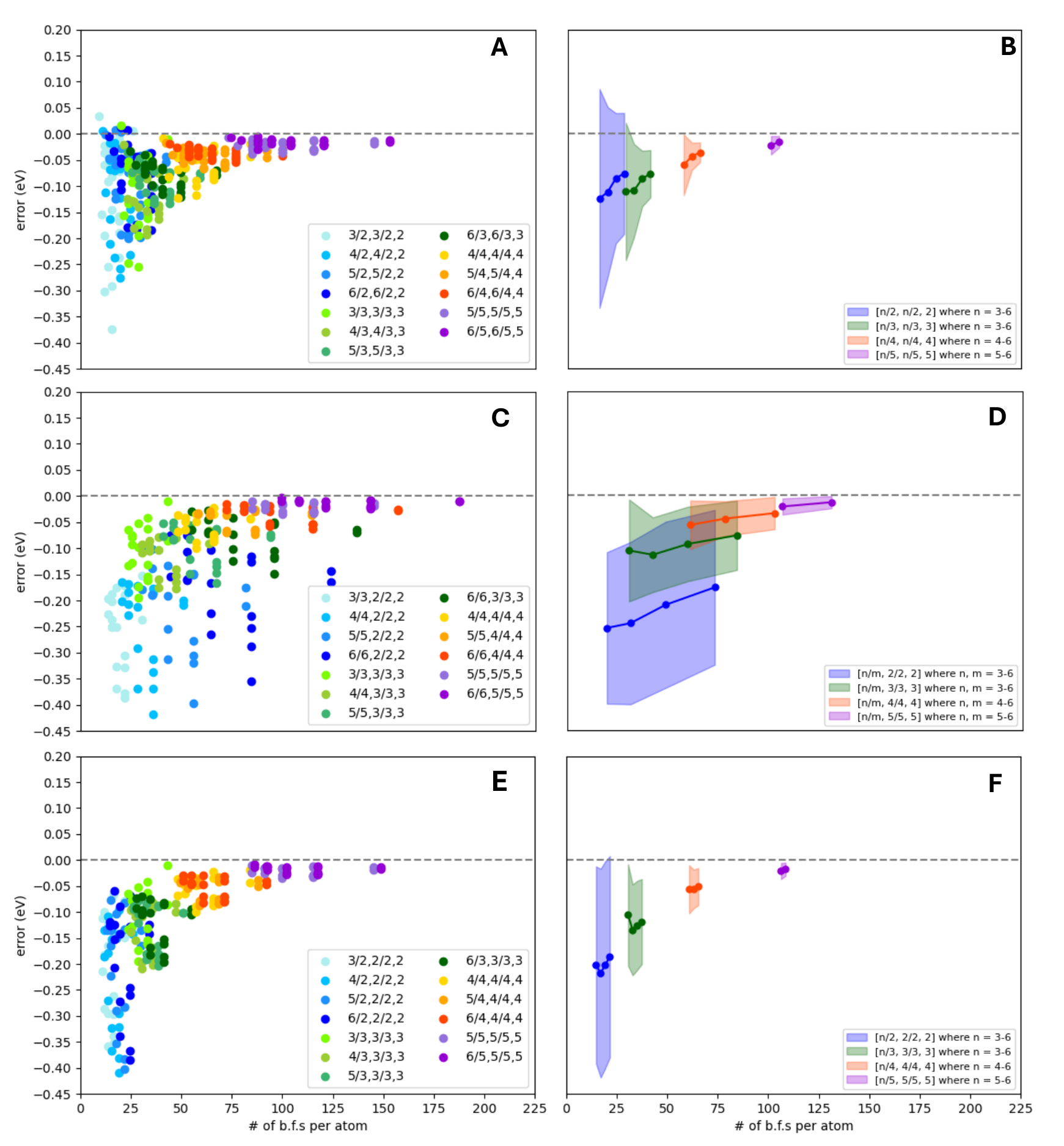}
\centering
\caption{
Errors in vertical IEs with respect to the estimated CBS limit for CVS-EOMIP-CCSD/AE with non-augmented basis sets. \textbf{A,C,E}: scatter plots of individual data . \textbf{B,D,F}: points represent average error for each basis set combination, and the shaded region indicates the 95\% confidence interval over the test set. \textbf{A,B}: hybrid basis sets applied to all heavy (non-hydrogen) atoms. \textbf{C,D}: standard basis sets, with different basis sets on active and non-active atoms (mixed basis sets). \textbf{E,F}: hybrid basis sets on active atoms, and (smaller) standard basis sets on other atoms (hybrid mixed basis sets).
}
\label{fig:NonAug}
\end{figure}

We begin by evaluating standard, hybrid, and mixed basis sets which do not include diffuse functions, but do include tight core functions. With these basis sets we also do not freeze any core electrons, even on inactive atoms. These results are summarized in Fig.~\ref{fig:NonAug}. 


\subsection{Hybrid Basis Sets}

Fig.~\ref{fig:NonAug}A,B depicts the individual and statistical errors for hybrid basis sets of varying quality. The basis sets $n/m$ $\equiv$ [$n$/$m$, $n$/$m$, $m$] are grouped by the cardinal number $m$, and the standard cc-pCV$m$Z basis sets are equivalent to the smallest basis set in each group ($n/m$), except for cc-pCVDZ which is not reported due to predictably and unacceptably large errors. The individual errors show a clear trend towards smaller errors for larger basis sets (either by increasing $n$ or $m$), and a consistent underestimation of the IEs. There is a significant overlap between the data points for $n/m$ with higher $n$ and $n$/$m+1$ for lower $n$ (compared to the corresponding $m$). This suggests that improved results can indeed be obtained either by increasing the size of the entire basis along the standard cc-pCV$n$Z hierarchy or by expanding only the $sp$ set. For example, results for $6/2$ are similar in both quality and basis set size to $3/3$. This result is also borne out in Fig.~\ref{fig:NonAug}B, where we clearly see that the number of b.f.s per atom and 95\% confidence interval of the error are very similar for those bases. From that subfigure we also observe that the errors in each $n/m$ sequence decrease with increasing $n$ (e.g. from $\pm\SI{0.210} {\eV}$ for $3/2$ to $\pm\SI{0.116}{\eV}$ for $6/2$ and from $\pm\SI{0.132}{\eV}$ for $3/3$ to $\pm\SI{0.045}{\eV}$ for $6/3$). 


Average errors also decrease with increasing $m$, but the reduction for the $n/2$ and $n/3$ series is rather moderate, and the reduction of variability is much more significant. The largest ``DZ'' set $6/2$ and smallest ``TZ'' set $3/3$, as noted above, essentially overlap in error and size. However, there is a significant gap between $6/3$ and $4/4$ and between $6/4$ and $5/5$. In each case, the smaller $6/m$ basis set achieves a similar range of errors as the larger $m+1/m+1$ set, but with a much smaller number of basis functions. This could result in significant savings for calculations targeting high basis set accuracy ($\sim\SI{0.1}{\eV}$) by using hybrid ``TZ'' basis sets, as high as perhaps $3.7\times$ based on the ratio of number of b.f.s per atom (57 for $4/4$ vs. 41 for $6/3$) and the $\mathcal{O}(V^4)$ scaling with basis set of CCSD.


\subsection{Mixed Basis Sets}

In Fig.~\ref{fig:NonAug}C,D we explore mixing different basis sets on active and inactive atoms, starting with mixtures of standard cc-pCV$n$Z basis sets. These mixed basis sets [$n$/$n$, $m$/$m$, $m$] are grouped by the $m$ cardinal number as above. In these results, we exclude systems with only one heavy atom since they do benefit from these mixed basis set schemes. The individual errors steadily decrease as the basis set size increases, rapidly w.r.t the cardinal number $m$, and more slowly with the cardinal number $n$. In comparison to the results from the hybrid basis set schemes (Fig.~\ref{fig:NonAug}A,B) the data points are not as tightly clustered. It is clear that the [$n$/$n$,$2$,$2$] basis set schemes, regardless of the value of $n$ from 3 up to 6, yield considerably underestimated IEs. Regarding the other schemes, it is clear that the estimation of IEs improves considerably as $m$ increases. Despite the small increase in accuracy on increasing $n$, the rapid increase in basis set size renders this strategy alone ineffective. For example, while [6/2, 6/2, 2] and [3/3, 3/3, 3] are similar in error (Fig.~\ref{fig:NonAug}A,B), [6/6, 2/2, 2] is substantially less accurate than [3/3, 3/3, 3] ($-0.176\pm\SI{0.148}{\eV}$ compared to $-0.105\pm\SI{0.098}{\eV}$, respectively), while requiring approximately $2.5\times$ as many b.f.s on average. Of course, the relative sizes of the basis sets will change (favoring [6/6, 2/2, 2]) for larger molecules, but the use of at least TZ-quality basis sets for the entire molecule is required to bring errors below 0.2 eV when using only standard basis sets.

In Fig.~\ref{fig:NonAug}E,F we investigate mixing different hybrid and standard basis sets on active and inactive atoms. A hybrid basis set is used on the active atom and standard basis sets are used on inactive atoms, i.e. [$n$/$m$, $m$/$m$, $m$]. Basis sets are grouped by $m$ as before. While individual points (panel E) are still highly spread along the $y$-axis, the grouping along the $x$-axis is much tigher, with visible gaps between different groups defined by differing $m$ cardinal numbers. However, the errors within and between each basis set group still closely follow those observed in Fig.~\ref{fig:NonAug}C,D. As before, increasing $m$ has a significant impact on accuracy, while increasing $n$ only has little to no effect. However, the use of hybrid basis sets significantly decreases the number of basis sets required without reducing accuracy.

The combination of these results points out an interesting finding: the basis sets on neighboring atoms are critical in reducing accuracy below the $\sim 0.3 eV$ level. This is illustrated, for example, in the comparison between increasing $n$ and $m$ in the [$n$/$n$, $m$/$m$, $m$] mixed standard basis sets, as well as in the mixed hybrid basis sets. Here, even using a 6Z basis set on the active atom yields large errors when only a DZ set is used on neighboring atoms. Partly, this may be due to an overall poorer description of the general molecular structure which, due to bonding and other delocalized interactions, requires sufficient basis set description beyond the active atom. However, the clear and significant decrease in error for the hybrid (non-mixed) [$n$/2, $n$/2, 2] series argues for a more specific effect. Here, the imbalance in the description of neighboring atoms, and particularly the poor description of polarization, should still negatively impact bonding and other interatomic interactions. Instead, the expanded $sp$ set alone, even on inactive atoms, improves errors to below 0.2 eV. We will revisit this effect in the next section.

\begin{figure}
\includegraphics[width=\textwidth]{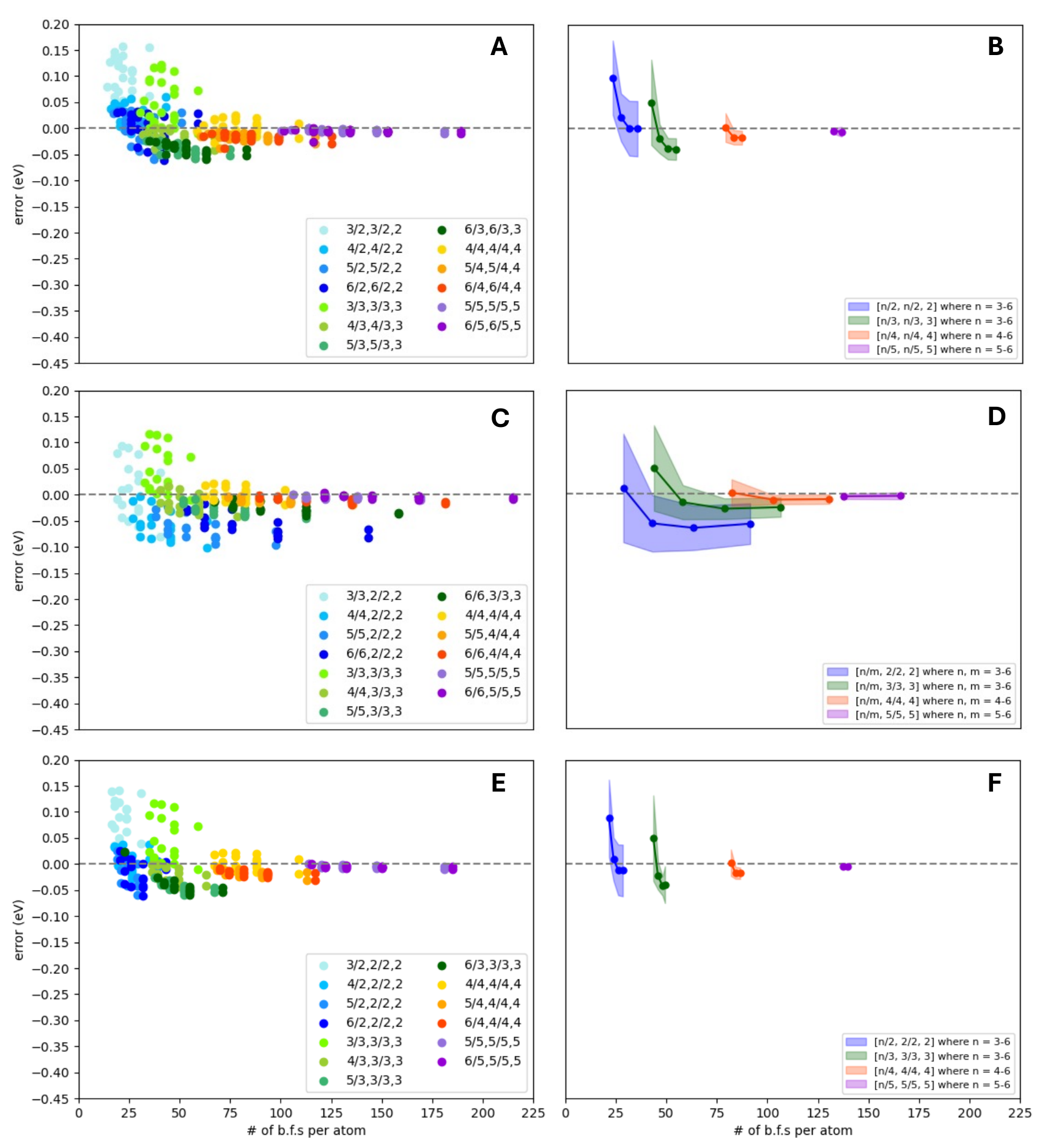}
\centering
\caption{
As in Fig.~\ref{fig:NonAug}, but for CVS-EOMIP-CCSD/AE with augmented basis sets. \textbf{A,B}: hybrid basis sets; \textbf{C,D}: standard mixed basis sets; \textbf{E,F}: hybrid mixed basis sets.
}
\label{fig:AUG}
\end{figure}

\subsection{Role of Diffuse Functions}


Fig.~\ref{fig:AUG}A,B depicts the individual and statistical errors for results computed using augmented hybrid basis sets aug-[$n$/$m$, $n$/$m$, $m$]. The data shows a consistent decrease in error as the basis set size increases, particularly with respect to $n$. Most such schemes are shown to yield individual errors within 0.1 eV of the estimated CBS limit. There is also remarkable overlap among the data for $n/m$ with large $n$ and $m+1/m+1$. From Fig.~\ref{fig:AUG}B we see the average errors in each $n/m$ sequence generally become smaller as $m$ increases, although the DZ series $n$/2 seems to benefit from fortuitous error cancellation. Interestingly, the average error of the augmented 6/2 basis set is now considerably less than for the slightly larger 3/3 ($\equiv$aug-cc-pCVTZ) basis set, although the 95\% CI is similar. In comparison to the non-augmented hybrid basis set schemes (Fig.~\ref{fig:NonAug}A,B), the addition of diffuse functions leads to significantly smaller average errors and improved convergence behavior with increasing basis set size. In most cases, the inclusion of diffuse functions reduces the average errors for the hybrid basis schemes by 70\%, while average basis set size increases by only 30\%. The 6/2 hybrid basis, with the inclusion of diffuse functions, now obtains a similar error as cc-pCVQZ, compared to the similarity of non-augmented 6/2 to cc-pCVTZ. The 40\% reduction in number of basis functions from cc-pCVQZ to aug-6/2 promises speedups of up to $4\times$ with similar basis set errors below 0.1 eV.



Fig.~\ref{fig:AUG}C,D presents the distribution of errors obtained for results computed using augmented standard mixed basis schemes, aug-[$n$/$n$, $m$/$m$, $m$]. Overall, the errors are very similar to the hybrid basis sets, except that many more basis functions are requires. Additionally, the aug-[$n$/$n$, 2/2, 2] series loses the fortuitous error cancellation observed in the hybrid case. The rapid growth in basis size with $n$ leads to basis sets with at least TZ quality for all atoms being the best compromise, although for larger molecules basis sets such as aug-[5/5, 2/2, 2] may be the most effective combination in this category. When considering mixed hybrid basis sets (Fig.~\ref{fig:AUG}E,F), we see almost exactly the same errors on a per-basis-set level as with non-mixed hybrid sets (panels A,B), along with somewhat reduced numbers of basis functions due to the use of standard aug-cc-pCVDZ basis sets on all other atoms. So far, these basis sets clearly provide the most effective compromise of efficiency (small basis size) and accuracy, although the accuracy achievable at the e.g. aug-[5/2, 2/2, 2] level already a significant improvement over standard cc-pCVDZ, cc-pCVTZ, and even cc-pCVQZ basis sets.

\begin{figure}
\includegraphics[width=\textwidth]{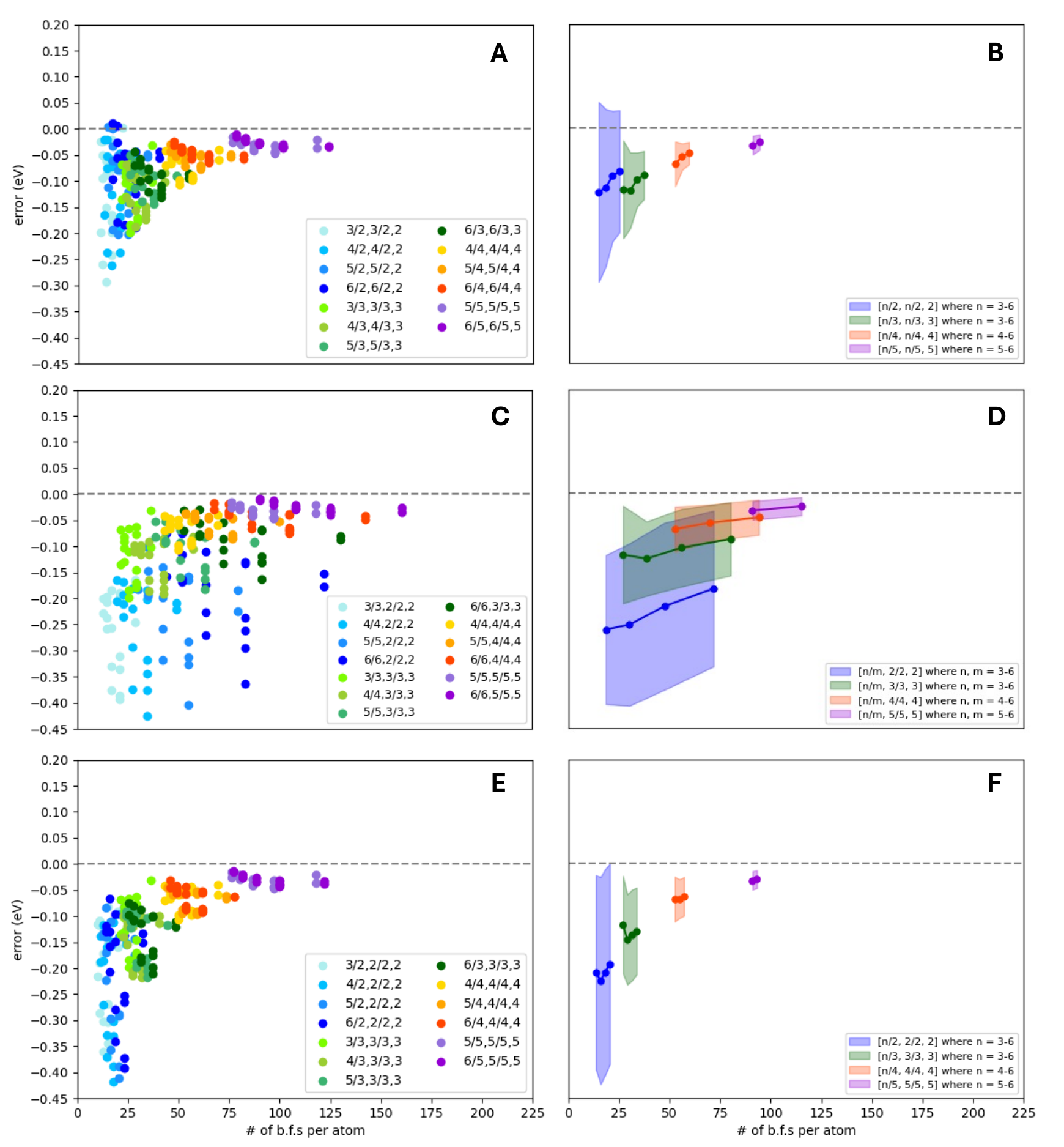}
\centering
\caption{
As in Fig.~\ref{fig:NonAug}, but for CVS-EOMIP-CCSD/FC with non-augmented basis sets. \textbf{A,B}: hybrid basis sets; \textbf{C,D}: standard mixed basis sets; \textbf{E,F}: hybrid mixed basis sets.
}
\label{fig:FC-NonAug}
\end{figure}

Clearly, the inclusion of diffuse functions significantly improves the accuracy of the calculated 1s IEs in our test set. The smaller but still significant improvement along the non-augmented [$n$/2, $n$/2, 2] series (Fig.~\ref{fig:NonAug}A,B) also indicates that including larger $sp$ sets on neighboring atoms has a beneficial effect. This then raises the question whether the effect of diffuse functions is more significant on the active atom or on inactive atoms. We performed additional tests with basis sets of the type [aug-$n$/$n$, $m$/$m$, $m$] and [aug-$n$/$m$, $m$/$m$, $m$] where diffuse functions are only included on the active atoms. Fig.~S1A-D in the Supplementary Information shows that this is clearly not an effective strategy, with errors as large as for non-augmented mixed basis sets (Fig.~\ref{fig:NonAug}C-F). Thus, the reduction in error stems from increased flexibility in the basis sets of neighboring atoms. In particular, the addition of diffuse functions on nearby atoms (either explicitly diffuse functions or additional uncontracted functions present in ``higher'' Dunning basis sets, which also tend to have smaller exponents), even if only of $s$ or $p$ type, is necessary to reach errors below 0.1 eV. This is likely due to better representation of polarization of the core orbital due to bonding interactions; in the active atom basis this requires functions of relatively large angular momentum, while low angular momentum functions on neighboring atoms are already suitably oriented.

\subsection{Frozen Core Approximation}

Here we explore the use of frozen core approximation to all hybrid and mixed basis sets, with and without diffuse functions included. In these calculations the core electrons of inactive atoms are frozen and the basis sets assigned to these atoms do not contain tight core functions (e.g. aug-cc-pVnZ, cc-pVnZ). In these results, we exclude systems with only one heavy atom. Fig.~\ref{fig:FC-NonAug}A-F show the individual and statistical errors for hybrid, standard mixed, and hybrid mixed basis sets with the frozen core (FC) approximation included on inactive atoms. We employ cc-pV$n$Z basis sets on these atoms rather than cc-pCV$n$Z as before. In comparison to Fig.~\ref{fig:NonAug}, we see similar slightly larger errors in most cases. For example, the hybrid $n$/$m$ series (panels A,B) average errors of $-0.1222\pm 0.0863$ eV for 3/2 to $-0.0822\pm 0.0586$ for 6/2, compared to $-0.1239\pm 0.1050$ eV and $-0.0761\pm 0.0580$ eV, respectively, without frozen core.

\begin{figure}
\includegraphics[width=\textwidth]{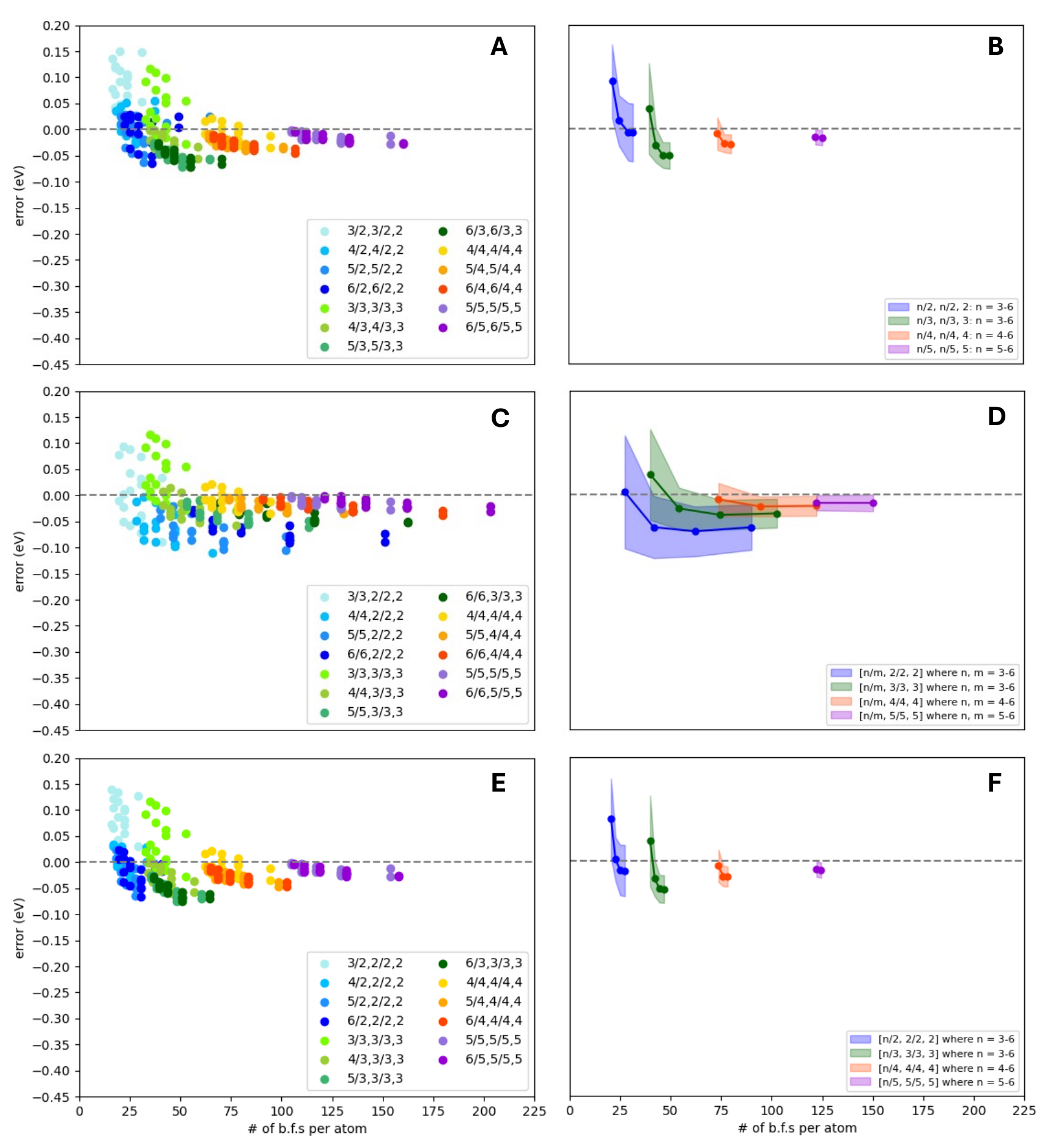}
\centering
\caption{
As in Fig.~\ref{fig:NonAug}, but for CVS-EOMIP-CCSD/FC with augmented basis sets. \textbf{A,B}: hybrid basis sets; \textbf{C,D}: standard mixed basis sets; \textbf{E,F}: hybrid mixed basis sets.
}
\label{fig:FC-Aug}
\end{figure}

Increases in average errors are slightly larger for TZ, QZ, and 5Z basis sets, converging to $\sim 12$ meV near the basis set limit. This small offset is expected due to the lack of core-valence correlation in inactive atoms, which has a measurable although weak effect on the core ionization energy. Note that the active atom has all electrons correlated in all cases. The 95\% confidence intervals experience the largest increase of 19 meV for 3/2 and 3/3 basis sets. The next largest increases in 95\% CIs are only 8 meV. The mixed basis set schemes, both with and without a hybrid basis set on the active atom, show the most regular behavior with respect to freezing core electrons. Average errors increase consistently by 6--7 meV for combinations with DZ bases on inactive atoms, and 11--12 meV for larger basis sets (in fact, this is a reduction in error compared to the FC basis set limit). 95\% CIs increase by at most 3 meV.

\begin{figure}
\includegraphics[width=\textwidth]{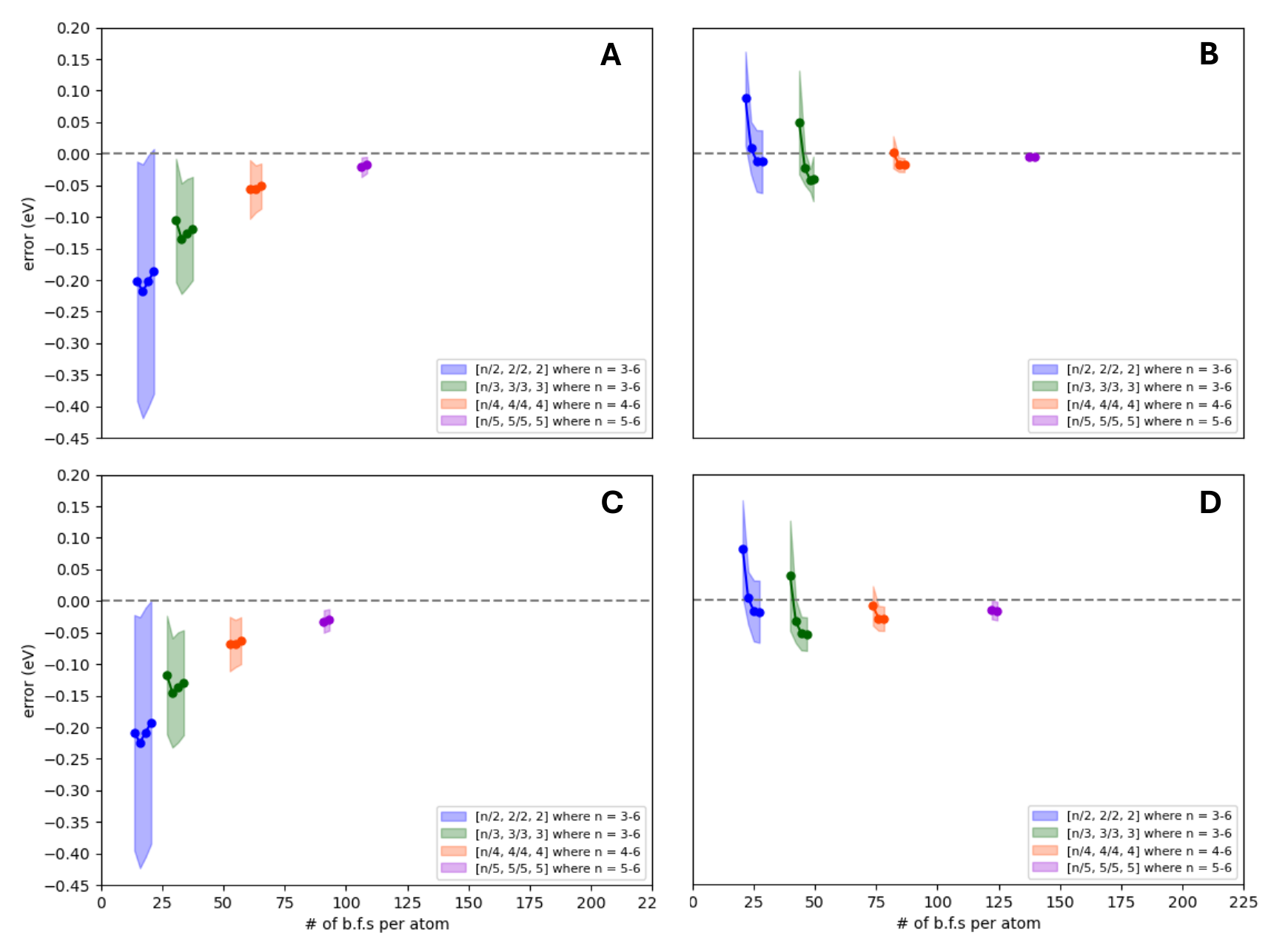}
\centering
\caption{
Comparison of panel \textbf{F} (hybrid mixed basis sets, 95\% confidence intervals of IE errors with respect to estimated CBS limit) from Figs.~\ref{fig:NonAug}--\ref{fig:FC-Aug}. \textbf{A}: all electron, non-augmented basis sets (Fig.~\ref{fig:NonAug}). \textbf{B}: all electron, augmented basis sets (Fig.~\ref{fig:AUG}). \textbf{C}: frozen core, non-augmented basis sets (Fig.~\ref{fig:FC-NonAug}). \textbf{D}: frozen core, augmented basis sets (Fig.~\ref{fig:FC-Aug}).
}
\label{fig:compare-mixed}
\end{figure}

The effect of the frozen core approximation with augmented basis sets is depicted in Fig.~\ref{fig:FC-Aug}. These results show very similar changes in errors compared to the augmented all-electron results (Fig.~\ref{fig:AUG}) as were observed in the non-augmented case. However, the changes are somewhat less systematic, particularly for mixed basis sets. Interestingly, the maximum increases in error are highly similar to those noted above: up to 14 meV increase in average error and 17 meV increase in 95\% CI (8 meV without the extremal case of [3/3, 2/2, 2]). Instead, the variability of the frozen core shifts tends to result in fortuitous error cancellation, leading in some cases to \emph{decreases} in average errors by 10--12 meV. The most direct comparisons can be made in Fig.~\ref{fig:compare-mixed}, where the hybrid mixed basis set results for all combinations of augmented/non-augmented and all electron/frozen core are presented together. Here, it is almost impossible to distinguish the additional errors due to freezing core electrons visually, while some slight decrease in the basis set size (particularly for larger basis sets) is notable due to the removal of tight core functions on inactive atoms.

\subsection{Comparison to Other Schemes}

\begin{figure}
  \includegraphics[width=\textwidth,height=19cm]{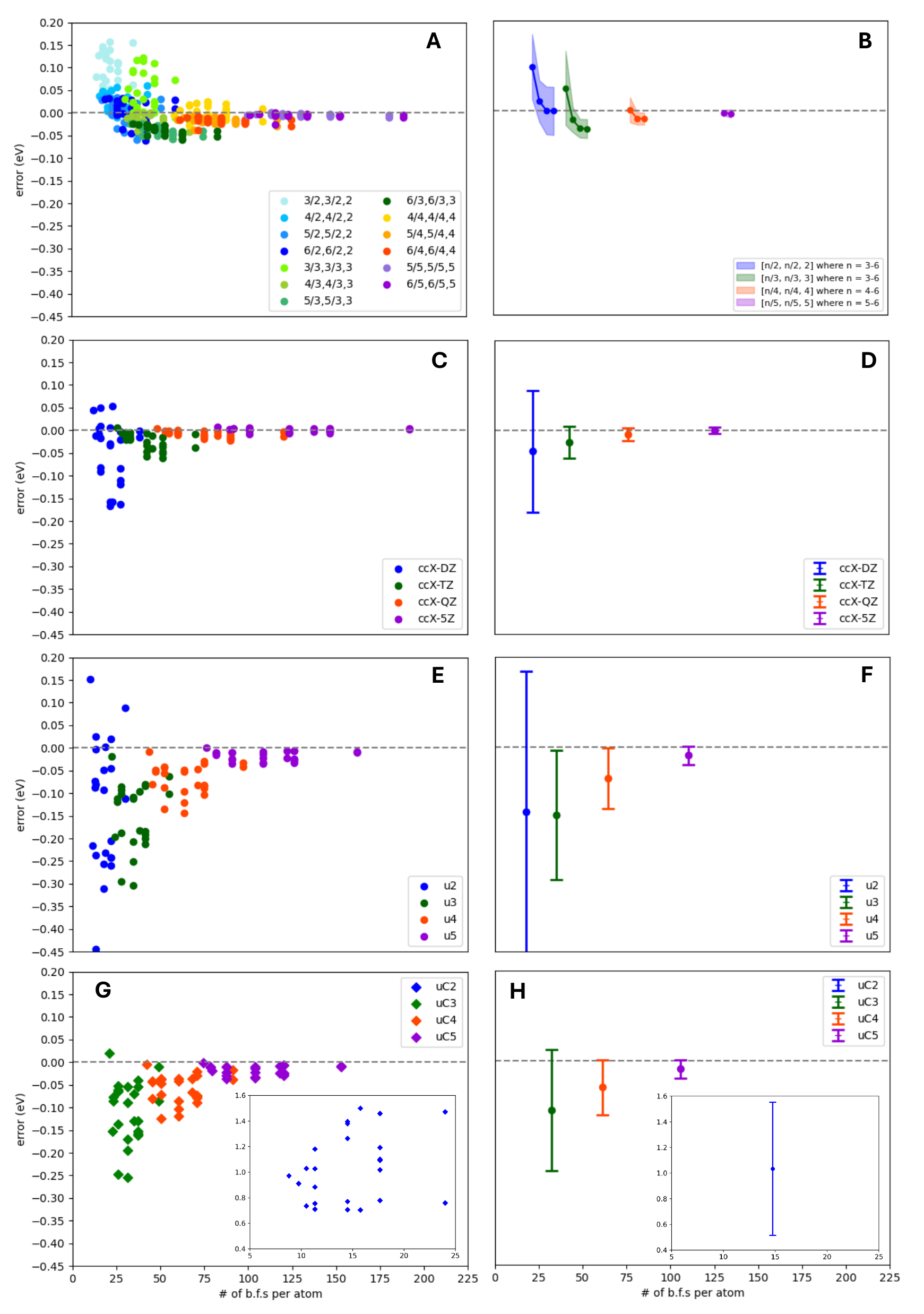}
  \caption{Comparison of IE errors w.r.t. estimated CBS limit for different basis sets, using CVS-EOMIP-CCSD/AE. \textbf{A,B}: augmented hybrid basis sets; \textbf{C,D}: ccX-$n$Z basis sets\cite{Ambroise2021}; \textbf{E,F}: uncontracted cc-pCV$n$Z basis sets; \textbf{G,H}: core-uncontracted cc-pCV$n$Z basis sets. In all cases the heavy atoms of the same molecule were described by equivalent basis sets and equivalent (contracted) standard cc-pV$n$Z basis sets were used for H atoms.}
  \label{fig:compare-outside}
\end{figure}

Finally, we compare our hybrid basis sets to existing schemes including the use of core- (uC$n$) or fully-uncontracted (u$n$) basis sets, as well the purpose-built ccX-$n$Z basis sets of Ambroise et al.\cite{Ambroise2021} As there is only scattered data for mixed basis set and mixed frozen core XPS calculations in the literature using these other approaches, we compare only calculations which employ the same basis set on all atoms (except hydrogen) and correlate all electrons; a full study of these additional effects for uncontracted and ccX-$n$Z bases is outside the present scope. Results are presented in Fig.~\ref{fig:compare-outside}. We include diffuse functions in our hybrid basis sets due to the clear benefit and small increase in basis set size, although such functions are not commonly used for XPS calculations in the literature. In comparison to uncontracted basis sets, we find a significant increase in accuracy at all basis set sizes. In particular, we notice very similarly behavior of uC$n$ and u$n$ basis sets of TZ and higher quality, but extremely poor results with uC2 compared to u2. Our hybrid ``DZ'' basis sets significantly outperform both of these uncontracted DZ sets. Both the DZ and TZ hybrid sets obtain errors better than uC4 or u4 at significantly reduced basis set size. In comparison to the ccX-$n$Z sets, we find generally similar or slightly smaller errors, for hyrid basis sets of similar or slightly larger size. The biggest gain in accuracy comes for the DZ hybrid sets, for example 5/2, in comparison to ccX-DZ. While 5/2 is still slightly less accurate than ccX-TZ, it is significantly more accurate than ccX-DZ ($0.0008\pm 0.0263$ eV and $-0.0459\pm 0.0671$, respectively), and only 39\% larger. In a mixed basis set calculation, this increase in size becomes essentially negligible.

\section{Conclusions}

In this work we investigate the basis set requirements for accurate computations of core-level ionization energies (IEs) using the CVS-EOM-IP-CCSD method across a set of 15 molecules, many with more than one heavy atom. From our results, we provide a detailed discussion on the use of standard and off-the-shelf hybrid and mixed basis set combinations which are constructed from Dunning’s augmented and non-augmented correlation-consistent basis sets. We also explored the use of the frozen core approximation for inactive atoms. From the IE values computed, we quantified IE errors for all our hybrid and mixed basis set schemes.

Our results show that hybrid and mixed basis sets can be a highly effective strategy for approaching the basis set limit in XPS calculations of organic molecules, at low computational cost. In particular, we find that diffuse functions should be included on all atoms (or potentially, on only the active and nearby atoms), and that a very compact mixture of high-$\zeta$ $sp$ functions (5Z or higher) with all other functions of DZ quality leads to reliably sub-0.1 eV basis set errors. We further find that applying the frozen core approximation to inactive atoms has an extremely minor effect on the computed IEs, typically below 10 meV. Some care must be taken in certain cases, chiefly when the active atom is equivalent to others by symmetry operations of the molecular point group. In this case, all equivalent atoms should be given the same basis set, have all electrons correlated, and by simultaneously included in the core-valence separation. This procedure may be necessary even in cases of accidental degeneracy (near-equivalency).

In comparison to existing basis sets approaches, we find that hybrid basis sets are significantly more effective that core- or fully-uncontracted Dunning basis sets, and are similarly effective compared to purpose-built basis sets such as ccX-$n$Z. We find the simplicity of the concept of hybrid basis sets highly attractive, and we hope that this approach will be readily extensible to a larger portion of the periodic table. For general calculations of organic K-edge ionization energies, we can now readily recommend the aug-[5/2, 2/2, 2]/FC scheme as our preferred method.

In future work, we also hope to validate the use of hybrid mixed basis sets for core-excited states relevant to x-ray absorption (XAS) and resonant inelastic x-ray scattering (RIXS) experiments.

\begin{acknowledgments}
This work was supported in part by the US National Science Foundation under grant CHE-2143725 and by the US Department of Energy under grant DE-SC0022893. Computational resources for this research were provided by SMU’s O’Donnell Data Science and Research Computing Institute.
\end{acknowledgments}

\section*{Conflict of Interest Statement}

The authors have no conflicts to disclose.

\section*{Supplementary Information}

Supplemental information files are available containing:
\begin{itemize}
\item Additional data and figures for partially-augmented basis sets (.pdf).
\item Computed CCSD total energies, ionization energies, ionization energy errors, and number of basis functions for all presented calculations. (.xlsx)
\end{itemize}

\bibliography{paper}

\begin{thebibliography}{68}%
\makeatletter
\providecommand \@ifxundefined [1]{%
 \@ifx{#1\undefined}
}%
\providecommand \@ifnum [1]{%
 \ifnum #1\expandafter \@firstoftwo
 \else \expandafter \@secondoftwo
 \fi
}%
\providecommand \@ifx [1]{%
 \ifx #1\expandafter \@firstoftwo
 \else \expandafter \@secondoftwo
 \fi
}%
\providecommand \natexlab [1]{#1}%
\providecommand \enquote  [1]{``#1''}%
\providecommand \bibnamefont  [1]{#1}%
\providecommand \bibfnamefont [1]{#1}%
\providecommand \citenamefont [1]{#1}%
\providecommand \href@noop [0]{\@secondoftwo}%
\providecommand \href [0]{\begingroup \@sanitize@url \@href}%
\providecommand \@href[1]{\@@startlink{#1}\@@href}%
\providecommand \@@href[1]{\endgroup#1\@@endlink}%
\providecommand \@sanitize@url [0]{\catcode `\\12\catcode `\$12\catcode
  `\&12\catcode `\#12\catcode `\^12\catcode `\_12\catcode `\%12\relax}%
\providecommand \@@startlink[1]{}%
\providecommand \@@endlink[0]{}%
\providecommand \url  [0]{\begingroup\@sanitize@url \@url }%
\providecommand \@url [1]{\endgroup\@href {#1}{\urlprefix }}%
\providecommand \urlprefix  [0]{URL }%
\providecommand \Eprint [0]{\href }%
\providecommand \doibase [0]{https://doi.org/}%
\providecommand \selectlanguage [0]{\@gobble}%
\providecommand \bibinfo  [0]{\@secondoftwo}%
\providecommand \bibfield  [0]{\@secondoftwo}%
\providecommand \translation [1]{[#1]}%
\providecommand \BibitemOpen [0]{}%
\providecommand \bibitemStop [0]{}%
\providecommand \bibitemNoStop [0]{.\EOS\space}%
\providecommand \EOS [0]{\spacefactor3000\relax}%
\providecommand \BibitemShut  [1]{\csname bibitem#1\endcsname}%
\let\auto@bib@innerbib\@empty
\bibitem [{\citenamefont {Bagus}(1965)}]{Bagus1965}%
  \BibitemOpen
  \bibfield  {author} {\bibinfo {author} {\bibfnamefont {P.~S.}\ \bibnamefont
  {Bagus}},\ }\bibfield  {title} {\enquote {\bibinfo {title}
  {Self-consistent-field wave functions for hole states of some ne-like and
  ar-like ions},}\ }\href {https://doi.org/10.1103/physrev.139.a619} {\bibfield
   {journal} {\bibinfo  {journal} {Physical Review}\ }\textbf {\bibinfo
  {volume} {139}},\ \bibinfo {pages} {A619–A634} (\bibinfo {year}
  {1965})}\BibitemShut {NoStop}%
\bibitem [{\citenamefont {Naves~de Brito}\ \emph {et~al.}(1991)\citenamefont
  {Naves~de Brito}, \citenamefont {Correia}, \citenamefont {Svensson},\ and\
  \citenamefont {Ågren}}]{NavesdeBrito1991}%
  \BibitemOpen
  \bibfield  {author} {\bibinfo {author} {\bibfnamefont {A.}~\bibnamefont
  {Naves~de Brito}}, \bibinfo {author} {\bibfnamefont {N.}~\bibnamefont
  {Correia}}, \bibinfo {author} {\bibfnamefont {S.}~\bibnamefont {Svensson}},\
  and\ \bibinfo {author} {\bibfnamefont {H.}~\bibnamefont {Ågren}},\
  }\bibfield  {title} {\enquote {\bibinfo {title} {A theoretical study of x-ray
  photoelectron spectra of model molecules for polymethylmethacrylate},}\
  }\href {https://doi.org/10.1063/1.460898} {\bibfield  {journal} {\bibinfo
  {journal} {The Journal of Chemical Physics}\ }\textbf {\bibinfo {volume}
  {95}},\ \bibinfo {pages} {2965--2974} (\bibinfo {year} {1991})}\BibitemShut
  {NoStop}%
\bibitem [{\citenamefont {Schmitt}\ and\ \citenamefont
  {Schirmer}(1992)}]{Schmitt1992}%
  \BibitemOpen
  \bibfield  {author} {\bibinfo {author} {\bibfnamefont {A.}~\bibnamefont
  {Schmitt}}\ and\ \bibinfo {author} {\bibfnamefont {J.}~\bibnamefont
  {Schirmer}},\ }\bibfield  {title} {\enquote {\bibinfo {title} {Molecular
  k-shell excitation spectra in the relaxed-core hartree-fock approximation},}\
  }\href {https://doi.org/10.1016/0301-0104(92)87124-r} {\bibfield  {journal}
  {\bibinfo  {journal} {Chemical Physics}\ }\textbf {\bibinfo {volume} {164}},\
  \bibinfo {pages} {1--9} (\bibinfo {year} {1992})}\BibitemShut {NoStop}%
\bibitem [{\citenamefont {Besley}, \citenamefont {Gilbert},\ and\ \citenamefont
  {Gill}(2009)}]{Besley2009}%
  \BibitemOpen
  \bibfield  {author} {\bibinfo {author} {\bibfnamefont {N.~A.}\ \bibnamefont
  {Besley}}, \bibinfo {author} {\bibfnamefont {A.~T.~B.}\ \bibnamefont
  {Gilbert}},\ and\ \bibinfo {author} {\bibfnamefont {P.~M.~W.}\ \bibnamefont
  {Gill}},\ }\bibfield  {title} {\enquote {\bibinfo {title}
  {Self-consistent-field calculations of core excited states},}\ }\href
  {https://doi.org/10.1063/1.3092928} {\bibfield  {journal} {\bibinfo
  {journal} {The Journal of Chemical Physics}\ }\textbf {\bibinfo {volume}
  {130}} (\bibinfo {year} {2009}),\ 10.1063/1.3092928}\BibitemShut {NoStop}%
\bibitem [{\citenamefont {Triguero}\ \emph {et~al.}(1999)\citenamefont
  {Triguero}, \citenamefont {Plashkevych}, \citenamefont {Pettersson},\ and\
  \citenamefont {Ågren}}]{Triguero1999}%
  \BibitemOpen
  \bibfield  {author} {\bibinfo {author} {\bibfnamefont {L.}~\bibnamefont
  {Triguero}}, \bibinfo {author} {\bibfnamefont {O.}~\bibnamefont
  {Plashkevych}}, \bibinfo {author} {\bibfnamefont {L.}~\bibnamefont
  {Pettersson}},\ and\ \bibinfo {author} {\bibfnamefont {H.}~\bibnamefont
  {Ågren}},\ }\bibfield  {title} {\enquote {\bibinfo {title} {Separate state
  vs. transition state kohn-sham calculations of x-ray photoelectron binding
  energies and chemical shifts},}\ }\href
  {https://doi.org/10.1016/s0368-2048(99)00008-0} {\bibfield  {journal}
  {\bibinfo  {journal} {Journal of Electron Spectroscopy and Related
  Phenomena}\ }\textbf {\bibinfo {volume} {104}},\ \bibinfo {pages} {195--207}
  (\bibinfo {year} {1999})}\BibitemShut {NoStop}%
\bibitem [{\citenamefont {Jensen}, \citenamefont {J{\o}rgensen},\ and\
  \citenamefont {Ågren}(1987)}]{Jensen1987}%
  \BibitemOpen
  \bibfield  {author} {\bibinfo {author} {\bibfnamefont {H.~J.~A.}\
  \bibnamefont {Jensen}}, \bibinfo {author} {\bibfnamefont {P.}~\bibnamefont
  {J{\o}rgensen}},\ and\ \bibinfo {author} {\bibfnamefont {H.}~\bibnamefont
  {Ågren}},\ }\bibfield  {title} {\enquote {\bibinfo {title} {Efficient
  optimization of large scale mcscf wave functions with a restricted step
  algorithm},}\ }\href {https://doi.org/10.1063/1.453590} {\bibfield  {journal}
  {\bibinfo  {journal} {The Journal of Chemical Physics}\ }\textbf {\bibinfo
  {volume} {87}},\ \bibinfo {pages} {451--466} (\bibinfo {year}
  {1987})}\BibitemShut {NoStop}%
\bibitem [{\citenamefont {Ågren}\ \emph {et~al.}(1994)\citenamefont {Ågren},
  \citenamefont {Carravetta}, \citenamefont {Vahtras},\ and\ \citenamefont
  {Pettersson}}]{Aagren1994}%
  \BibitemOpen
  \bibfield  {author} {\bibinfo {author} {\bibfnamefont {H.}~\bibnamefont
  {Ågren}}, \bibinfo {author} {\bibfnamefont {V.}~\bibnamefont {Carravetta}},
  \bibinfo {author} {\bibfnamefont {O.}~\bibnamefont {Vahtras}},\ and\ \bibinfo
  {author} {\bibfnamefont {L.~G.}\ \bibnamefont {Pettersson}},\ }\bibfield
  {title} {\enquote {\bibinfo {title} {Direct, atomic orbital, static exchange
  calculations of photoabsorption spectra of large molecules and clusters},}\
  }\href {https://doi.org/10.1016/0009-2614(94)00318-1} {\bibfield  {journal}
  {\bibinfo  {journal} {Chemical Physics Letters}\ }\textbf {\bibinfo {volume}
  {222}},\ \bibinfo {pages} {75--81} (\bibinfo {year} {1994})}\BibitemShut
  {NoStop}%
\bibitem [{\citenamefont {Bagus}, \citenamefont {Sousa},\ and\ \citenamefont
  {Illas}(2016)}]{Bagus2016}%
  \BibitemOpen
  \bibfield  {author} {\bibinfo {author} {\bibfnamefont {P.~S.}\ \bibnamefont
  {Bagus}}, \bibinfo {author} {\bibfnamefont {C.}~\bibnamefont {Sousa}},\ and\
  \bibinfo {author} {\bibfnamefont {F.}~\bibnamefont {Illas}},\ }\bibfield
  {title} {\enquote {\bibinfo {title} {Consequences of electron correlation for
  xps binding energies: Representative case for c(1s) and o(1s) xps of co},}\
  }\href {https://doi.org/10.1063/1.4964320} {\bibfield  {journal} {\bibinfo
  {journal} {The Journal of Chemical Physics}\ }\textbf {\bibinfo {volume}
  {145}} (\bibinfo {year} {2016}),\ 10.1063/1.4964320}\BibitemShut {NoStop}%
\bibitem [{\citenamefont {Verma}, \citenamefont {Derricotte},\ and\
  \citenamefont {Evangelista}(2015)}]{Verma2015}%
  \BibitemOpen
  \bibfield  {author} {\bibinfo {author} {\bibfnamefont {P.}~\bibnamefont
  {Verma}}, \bibinfo {author} {\bibfnamefont {W.~D.}\ \bibnamefont
  {Derricotte}},\ and\ \bibinfo {author} {\bibfnamefont {F.~A.}\ \bibnamefont
  {Evangelista}},\ }\bibfield  {title} {\enquote {\bibinfo {title} {Predicting
  near edge x-ray absorption spectra with the spin-free exact-two-component
  hamiltonian and orthogonality constrained density functional theory},}\
  }\href {https://doi.org/10.1021/acs.jctc.5b00817} {\bibfield  {journal}
  {\bibinfo  {journal} {Journal of Chemical Theory and Computation}\ }\textbf
  {\bibinfo {volume} {12}},\ \bibinfo {pages} {144--156} (\bibinfo {year}
  {2015})}\BibitemShut {NoStop}%
\bibitem [{\citenamefont {Ekström}, \citenamefont {Norman},\ and\
  \citenamefont {Carravetta}(2006)}]{Ekstroem2006}%
  \BibitemOpen
  \bibfield  {author} {\bibinfo {author} {\bibfnamefont {U.}~\bibnamefont
  {Ekström}}, \bibinfo {author} {\bibfnamefont {P.}~\bibnamefont {Norman}},\
  and\ \bibinfo {author} {\bibfnamefont {V.}~\bibnamefont {Carravetta}},\
  }\bibfield  {title} {\enquote {\bibinfo {title} {Relativistic four-component
  static-exchange approximation for core-excitation processes in molecules},}\
  }\href {https://doi.org/10.1103/physreva.73.022501} {\bibfield  {journal}
  {\bibinfo  {journal} {Physical Review A}\ }\textbf {\bibinfo {volume} {73}}
  (\bibinfo {year} {2006}),\ 10.1103/physreva.73.022501}\BibitemShut {NoStop}%
\bibitem [{\citenamefont {Oosterbaan}, \citenamefont {White},\ and\
  \citenamefont {Head-Gordon}(2018)}]{Oosterbaan2018}%
  \BibitemOpen
  \bibfield  {author} {\bibinfo {author} {\bibfnamefont {K.~J.}\ \bibnamefont
  {Oosterbaan}}, \bibinfo {author} {\bibfnamefont {A.~F.}\ \bibnamefont
  {White}},\ and\ \bibinfo {author} {\bibfnamefont {M.}~\bibnamefont
  {Head-Gordon}},\ }\bibfield  {title} {\enquote {\bibinfo {title}
  {Non-orthogonal configuration interaction with single substitutions for the
  calculation of core-excited states},}\ }\href
  {https://doi.org/10.1063/1.5023051} {\bibfield  {journal} {\bibinfo
  {journal} {The Journal of Chemical Physics}\ }\textbf {\bibinfo {volume}
  {149}} (\bibinfo {year} {2018}),\ 10.1063/1.5023051}\BibitemShut {NoStop}%
\bibitem [{\citenamefont {Maganas}, \citenamefont {DeBeer},\ and\ \citenamefont
  {Neese}(2014)}]{Maganas2014}%
  \BibitemOpen
  \bibfield  {author} {\bibinfo {author} {\bibfnamefont {D.}~\bibnamefont
  {Maganas}}, \bibinfo {author} {\bibfnamefont {S.}~\bibnamefont {DeBeer}},\
  and\ \bibinfo {author} {\bibfnamefont {F.}~\bibnamefont {Neese}},\ }\bibfield
   {title} {\enquote {\bibinfo {title} {Restricted open-shell configuration
  interaction cluster calculations of the l-edge x-ray absorption study of tio2
  and caf2 solids},}\ }\href {https://doi.org/10.1021/ic500197v} {\bibfield
  {journal} {\bibinfo  {journal} {Inorganic Chemistry}\ }\textbf {\bibinfo
  {volume} {53}},\ \bibinfo {pages} {6374--6385} (\bibinfo {year}
  {2014})}\BibitemShut {NoStop}%
\bibitem [{\citenamefont {Toffoli}\ and\ \citenamefont
  {Decleva}(2016)}]{Toffoli2016}%
  \BibitemOpen
  \bibfield  {author} {\bibinfo {author} {\bibfnamefont {D.}~\bibnamefont
  {Toffoli}}\ and\ \bibinfo {author} {\bibfnamefont {P.}~\bibnamefont
  {Decleva}},\ }\bibfield  {title} {\enquote {\bibinfo {title} {A multichannel
  least-squares b-spline approach to molecular photoionization: Theory,
  implementation, and applications within the configuration–interaction
  singles approximation},}\ }\href {https://doi.org/10.1021/acs.jctc.6b00627}
  {\bibfield  {journal} {\bibinfo  {journal} {Journal of Chemical Theory and
  Computation}\ }\textbf {\bibinfo {volume} {12}},\ \bibinfo {pages}
  {4996--5008} (\bibinfo {year} {2016})}\BibitemShut {NoStop}%
\bibitem [{\citenamefont {Ehlert}\ and\ \citenamefont
  {Klamroth}(2016)}]{Ehlert2016}%
  \BibitemOpen
  \bibfield  {author} {\bibinfo {author} {\bibfnamefont {C.}~\bibnamefont
  {Ehlert}}\ and\ \bibinfo {author} {\bibfnamefont {T.}~\bibnamefont
  {Klamroth}},\ }\bibfield  {title} {\enquote {\bibinfo {title} {The quest for
  best suited references for configuration interaction singles calculations of
  core excited states},}\ }\href {https://doi.org/10.1002/jcc.24531} {\bibfield
   {journal} {\bibinfo  {journal} {Journal of Computational Chemistry}\
  }\textbf {\bibinfo {volume} {38}},\ \bibinfo {pages} {116--126} (\bibinfo
  {year} {2016})}\BibitemShut {NoStop}%
\bibitem [{\citenamefont {Asmuruf}\ and\ \citenamefont
  {Besley}(2008)}]{Asmuruf2008}%
  \BibitemOpen
  \bibfield  {author} {\bibinfo {author} {\bibfnamefont {F.~A.}\ \bibnamefont
  {Asmuruf}}\ and\ \bibinfo {author} {\bibfnamefont {N.~A.}\ \bibnamefont
  {Besley}},\ }\bibfield  {title} {\enquote {\bibinfo {title} {Calculation of
  near-edge x-ray absorption fine structure with the cis(d) method},}\ }\href
  {https://doi.org/10.1016/j.cplett.2008.08.054} {\bibfield  {journal}
  {\bibinfo  {journal} {Chemical Physics Letters}\ }\textbf {\bibinfo {volume}
  {463}},\ \bibinfo {pages} {267--271} (\bibinfo {year} {2008})}\BibitemShut
  {NoStop}%
\bibitem [{\citenamefont {DeBeer~George}, \citenamefont {Petrenko},\ and\
  \citenamefont {Neese}(2008)}]{DeBeerGeorge2008}%
  \BibitemOpen
  \bibfield  {author} {\bibinfo {author} {\bibfnamefont {S.}~\bibnamefont
  {DeBeer~George}}, \bibinfo {author} {\bibfnamefont {T.}~\bibnamefont
  {Petrenko}},\ and\ \bibinfo {author} {\bibfnamefont {F.}~\bibnamefont
  {Neese}},\ }\bibfield  {title} {\enquote {\bibinfo {title} {Prediction of
  iron k-edge absorption spectra using time-dependent density functional
  theory},}\ }\href {https://doi.org/10.1021/jp803174m} {\bibfield  {journal}
  {\bibinfo  {journal} {The Journal of Physical Chemistry A}\ }\textbf
  {\bibinfo {volume} {112}},\ \bibinfo {pages} {12936--12943} (\bibinfo {year}
  {2008})}\BibitemShut {NoStop}%
\bibitem [{\citenamefont {Vinson}\ \emph {et~al.}(2011)\citenamefont {Vinson},
  \citenamefont {Rehr}, \citenamefont {Kas},\ and\ \citenamefont
  {Shirley}}]{Vinson2011}%
  \BibitemOpen
  \bibfield  {author} {\bibinfo {author} {\bibfnamefont {J.}~\bibnamefont
  {Vinson}}, \bibinfo {author} {\bibfnamefont {J.~J.}\ \bibnamefont {Rehr}},
  \bibinfo {author} {\bibfnamefont {J.~J.}\ \bibnamefont {Kas}},\ and\ \bibinfo
  {author} {\bibfnamefont {E.~L.}\ \bibnamefont {Shirley}},\ }\bibfield
  {title} {\enquote {\bibinfo {title} {Bethe-salpeter equation calculations of
  core excitation spectra},}\ }\href
  {https://doi.org/10.1103/physrevb.83.115106} {\bibfield  {journal} {\bibinfo
  {journal} {Physical Review B}\ }\textbf {\bibinfo {volume} {83}} (\bibinfo
  {year} {2011}),\ 10.1103/physrevb.83.115106}\BibitemShut {NoStop}%
\bibitem [{\citenamefont {Barth}\ and\ \citenamefont
  {Schirmer}(1985)}]{Barth1985}%
  \BibitemOpen
  \bibfield  {author} {\bibinfo {author} {\bibfnamefont {A.}~\bibnamefont
  {Barth}}\ and\ \bibinfo {author} {\bibfnamefont {J.}~\bibnamefont
  {Schirmer}},\ }\bibfield  {title} {\enquote {\bibinfo {title} {Theoretical
  core-level excitation spectra of n2and co by a new polarisation propagator
  method},}\ }\href {https://doi.org/10.1088/0022-3700/18/5/008} {\bibfield
  {journal} {\bibinfo  {journal} {Journal of Physics B: Atomic and Molecular
  Physics}\ }\textbf {\bibinfo {volume} {18}},\ \bibinfo {pages} {867--885}
  (\bibinfo {year} {1985})}\BibitemShut {NoStop}%
\bibitem [{\citenamefont {Barth}\ and\ \citenamefont
  {Cederbaum}(1981)}]{Barth1981}%
  \BibitemOpen
  \bibfield  {author} {\bibinfo {author} {\bibfnamefont {A.}~\bibnamefont
  {Barth}}\ and\ \bibinfo {author} {\bibfnamefont {L.~S.}\ \bibnamefont
  {Cederbaum}},\ }\bibfield  {title} {\enquote {\bibinfo {title} {Many-body
  theory of core-valence excitations},}\ }\href
  {https://doi.org/10.1103/physreva.23.1038} {\bibfield  {journal} {\bibinfo
  {journal} {Physical Review A}\ }\textbf {\bibinfo {volume} {23}},\ \bibinfo
  {pages} {1038--1061} (\bibinfo {year} {1981})}\BibitemShut {NoStop}%
\bibitem [{\citenamefont {Trofimov}\ \emph {et~al.}(2000)\citenamefont
  {Trofimov}, \citenamefont {Moskovskaya}, \citenamefont {Gromov},
  \citenamefont {Vitkovskaya},\ and\ \citenamefont {Schirmer}}]{Trofimov2000}%
  \BibitemOpen
  \bibfield  {author} {\bibinfo {author} {\bibfnamefont {A.~B.}\ \bibnamefont
  {Trofimov}}, \bibinfo {author} {\bibfnamefont {T.~E.}\ \bibnamefont
  {Moskovskaya}}, \bibinfo {author} {\bibfnamefont {E.~V.}\ \bibnamefont
  {Gromov}}, \bibinfo {author} {\bibfnamefont {N.~M.}\ \bibnamefont
  {Vitkovskaya}},\ and\ \bibinfo {author} {\bibfnamefont {J.}~\bibnamefont
  {Schirmer}},\ }\bibfield  {title} {\enquote {\bibinfo {title} {Core-level
  electronic spectra in adc(2) approximation for polarization propagator:
  Carbon monoxide and nitrogen molecules},}\ }\href
  {https://doi.org/10.1007/bf02742009} {\bibfield  {journal} {\bibinfo
  {journal} {Journal of Structural Chemistry}\ }\textbf {\bibinfo {volume}
  {41}},\ \bibinfo {pages} {483--494} (\bibinfo {year} {2000})}\BibitemShut
  {NoStop}%
\bibitem [{\citenamefont {Wenzel}\ \emph {et~al.}(2015)\citenamefont {Wenzel},
  \citenamefont {Holzer}, \citenamefont {Wormit},\ and\ \citenamefont
  {Dreuw}}]{Wenzel2015}%
  \BibitemOpen
  \bibfield  {author} {\bibinfo {author} {\bibfnamefont {J.}~\bibnamefont
  {Wenzel}}, \bibinfo {author} {\bibfnamefont {A.}~\bibnamefont {Holzer}},
  \bibinfo {author} {\bibfnamefont {M.}~\bibnamefont {Wormit}},\ and\ \bibinfo
  {author} {\bibfnamefont {A.}~\bibnamefont {Dreuw}},\ }\bibfield  {title}
  {\enquote {\bibinfo {title} {Analysis and comparison of cvs-adc approaches up
  to third order for the calculation of core-excited states},}\ }\href
  {https://doi.org/10.1063/1.4921841} {\bibfield  {journal} {\bibinfo
  {journal} {The Journal of Chemical Physics}\ }\textbf {\bibinfo {volume}
  {142}} (\bibinfo {year} {2015}),\ 10.1063/1.4921841}\BibitemShut {NoStop}%
\bibitem [{\citenamefont {Southworth}\ \emph {et~al.}(2015)\citenamefont
  {Southworth}, \citenamefont {Wehlitz}, \citenamefont {Picón}, \citenamefont
  {Lehmann}, \citenamefont {Cheng},\ and\ \citenamefont
  {Stanton}}]{Southworth2015}%
  \BibitemOpen
  \bibfield  {author} {\bibinfo {author} {\bibfnamefont {S.~H.}\ \bibnamefont
  {Southworth}}, \bibinfo {author} {\bibfnamefont {R.}~\bibnamefont {Wehlitz}},
  \bibinfo {author} {\bibfnamefont {A.}~\bibnamefont {Picón}}, \bibinfo
  {author} {\bibfnamefont {C.~S.}\ \bibnamefont {Lehmann}}, \bibinfo {author}
  {\bibfnamefont {L.}~\bibnamefont {Cheng}},\ and\ \bibinfo {author}
  {\bibfnamefont {J.~F.}\ \bibnamefont {Stanton}},\ }\bibfield  {title}
  {\enquote {\bibinfo {title} {Inner-shell photoionization and core-hole decay
  of xe and xef2},}\ }\href {https://doi.org/10.1063/1.4922208} {\bibfield
  {journal} {\bibinfo  {journal} {The Journal of Chemical Physics}\ }\textbf
  {\bibinfo {volume} {142}} (\bibinfo {year} {2015}),\
  10.1063/1.4922208}\BibitemShut {NoStop}%
\bibitem [{\citenamefont {Peng}\ \emph {et~al.}(2015)\citenamefont {Peng},
  \citenamefont {Lestrange}, \citenamefont {Goings}, \citenamefont {Caricato},\
  and\ \citenamefont {Li}}]{Peng2015}%
  \BibitemOpen
  \bibfield  {author} {\bibinfo {author} {\bibfnamefont {B.}~\bibnamefont
  {Peng}}, \bibinfo {author} {\bibfnamefont {P.~J.}\ \bibnamefont {Lestrange}},
  \bibinfo {author} {\bibfnamefont {J.~J.}\ \bibnamefont {Goings}}, \bibinfo
  {author} {\bibfnamefont {M.}~\bibnamefont {Caricato}},\ and\ \bibinfo
  {author} {\bibfnamefont {X.}~\bibnamefont {Li}},\ }\bibfield  {title}
  {\enquote {\bibinfo {title} {Energy-specific equation-of-motion
  coupled-cluster methods for high-energy excited states: Application to k-edge
  x-ray absorption spectroscopy},}\ }\href
  {https://doi.org/10.1021/acs.jctc.5b00459} {\bibfield  {journal} {\bibinfo
  {journal} {Journal of Chemical Theory and Computation}\ }\textbf {\bibinfo
  {volume} {11}},\ \bibinfo {pages} {4146--4153} (\bibinfo {year}
  {2015})}\BibitemShut {NoStop}%
\bibitem [{\citenamefont {Coriani}\ \emph {et~al.}(2012)\citenamefont
  {Coriani}, \citenamefont {Christiansen}, \citenamefont {Fransson},\ and\
  \citenamefont {Norman}}]{Coriani2012}%
  \BibitemOpen
  \bibfield  {author} {\bibinfo {author} {\bibfnamefont {S.}~\bibnamefont
  {Coriani}}, \bibinfo {author} {\bibfnamefont {O.}~\bibnamefont
  {Christiansen}}, \bibinfo {author} {\bibfnamefont {T.}~\bibnamefont
  {Fransson}},\ and\ \bibinfo {author} {\bibfnamefont {P.}~\bibnamefont
  {Norman}},\ }\bibfield  {title} {\enquote {\bibinfo {title} {Coupled-cluster
  response theory for near-edge x-ray-absorption fine structure of atoms and
  molecules},}\ }\href {https://doi.org/10.1103/physreva.85.022507} {\bibfield
  {journal} {\bibinfo  {journal} {Physical Review A}\ }\textbf {\bibinfo
  {volume} {85}} (\bibinfo {year} {2012}),\
  10.1103/physreva.85.022507}\BibitemShut {NoStop}%
\bibitem [{\citenamefont {Coriani}\ and\ \citenamefont
  {Koch}(2015{\natexlab{a}})}]{Coriani2015}%
  \BibitemOpen
  \bibfield  {author} {\bibinfo {author} {\bibfnamefont {S.}~\bibnamefont
  {Coriani}}\ and\ \bibinfo {author} {\bibfnamefont {H.}~\bibnamefont {Koch}},\
  }\bibfield  {title} {\enquote {\bibinfo {title} {Communication: X-ray
  absorption spectra and core-ionization potentials within a core-valence
  separated coupled cluster framework},}\ }\href
  {https://doi.org/10.1063/1.4935712} {\bibfield  {journal} {\bibinfo
  {journal} {The Journal of Chemical Physics}\ }\textbf {\bibinfo {volume}
  {143}} (\bibinfo {year} {2015}{\natexlab{a}}),\
  10.1063/1.4935712}\BibitemShut {NoStop}%
\bibitem [{\citenamefont {Datta}\ and\ \citenamefont
  {Mukherjee}(2009)}]{Datta2009}%
  \BibitemOpen
  \bibfield  {author} {\bibinfo {author} {\bibfnamefont {D.}~\bibnamefont
  {Datta}}\ and\ \bibinfo {author} {\bibfnamefont {D.}~\bibnamefont
  {Mukherjee}},\ }\bibfield  {title} {\enquote {\bibinfo {title} {An explicitly
  spin-free compact open-shell coupled cluster theory using a multireference
  combinatoric exponential ansatz: Formal development and pilot
  applications},}\ }\href {https://doi.org/10.1063/1.3185356} {\bibfield
  {journal} {\bibinfo  {journal} {The Journal of Chemical Physics}\ }\textbf
  {\bibinfo {volume} {131}} (\bibinfo {year} {2009}),\
  10.1063/1.3185356}\BibitemShut {NoStop}%
\bibitem [{\citenamefont {Brabec}\ \emph {et~al.}(2012)\citenamefont {Brabec},
  \citenamefont {Bhaskaran-Nair}, \citenamefont {Govind}, \citenamefont
  {Pittner},\ and\ \citenamefont {Kowalski}}]{Brabec2012}%
  \BibitemOpen
  \bibfield  {author} {\bibinfo {author} {\bibfnamefont {J.}~\bibnamefont
  {Brabec}}, \bibinfo {author} {\bibfnamefont {K.}~\bibnamefont
  {Bhaskaran-Nair}}, \bibinfo {author} {\bibfnamefont {N.}~\bibnamefont
  {Govind}}, \bibinfo {author} {\bibfnamefont {J.}~\bibnamefont {Pittner}},\
  and\ \bibinfo {author} {\bibfnamefont {K.}~\bibnamefont {Kowalski}},\
  }\bibfield  {title} {\enquote {\bibinfo {title} {Communication: Application
  of state-specific multireference coupled cluster methods to core-level
  excitations},}\ }\href {https://doi.org/10.1063/1.4764355} {\bibfield
  {journal} {\bibinfo  {journal} {The Journal of Chemical Physics}\ }\textbf
  {\bibinfo {volume} {137}} (\bibinfo {year} {2012}),\
  10.1063/1.4764355}\BibitemShut {NoStop}%
\bibitem [{\citenamefont {Sen}, \citenamefont {Shee},\ and\ \citenamefont
  {Mukherjee}(2013)}]{Sen2013}%
  \BibitemOpen
  \bibfield  {author} {\bibinfo {author} {\bibfnamefont {S.}~\bibnamefont
  {Sen}}, \bibinfo {author} {\bibfnamefont {A.}~\bibnamefont {Shee}},\ and\
  \bibinfo {author} {\bibfnamefont {D.}~\bibnamefont {Mukherjee}},\ }\bibfield
  {title} {\enquote {\bibinfo {title} {A study of the ionisation and excitation
  energies of core electrons using a unitary group adapted state universal
  approach},}\ }\href {https://doi.org/10.1080/00268976.2013.802384} {\bibfield
   {journal} {\bibinfo  {journal} {Molecular Physics}\ }\textbf {\bibinfo
  {volume} {111}},\ \bibinfo {pages} {2625--2639} (\bibinfo {year}
  {2013})}\BibitemShut {NoStop}%
\bibitem [{\citenamefont {Dutta}\ \emph {et~al.}(2014)\citenamefont {Dutta},
  \citenamefont {Gupta}, \citenamefont {Vaval},\ and\ \citenamefont
  {Pal}}]{Dutta2014}%
  \BibitemOpen
  \bibfield  {author} {\bibinfo {author} {\bibfnamefont {A.~K.}\ \bibnamefont
  {Dutta}}, \bibinfo {author} {\bibfnamefont {J.}~\bibnamefont {Gupta}},
  \bibinfo {author} {\bibfnamefont {N.}~\bibnamefont {Vaval}},\ and\ \bibinfo
  {author} {\bibfnamefont {S.}~\bibnamefont {Pal}},\ }\bibfield  {title}
  {\enquote {\bibinfo {title} {Intermediate hamiltonian fock space
  multireference coupled cluster approach to core excitation spectra},}\ }\href
  {https://doi.org/10.1021/ct500285e} {\bibfield  {journal} {\bibinfo
  {journal} {Journal of Chemical Theory and Computation}\ }\textbf {\bibinfo
  {volume} {10}},\ \bibinfo {pages} {3656--3668} (\bibinfo {year}
  {2014})}\BibitemShut {NoStop}%
\bibitem [{\citenamefont {Emrich}(1981)}]{Emrich1981}%
  \BibitemOpen
  \bibfield  {author} {\bibinfo {author} {\bibfnamefont {K.}~\bibnamefont
  {Emrich}},\ }\bibfield  {title} {\enquote {\bibinfo {title} {An extension of
  the coupled cluster formalism to excited states (i)},}\ }\href
  {https://doi.org/10.1016/0375-9474(81)90179-2} {\bibfield  {journal}
  {\bibinfo  {journal} {Nuclear Physics A}\ }\textbf {\bibinfo {volume}
  {351}},\ \bibinfo {pages} {379--396} (\bibinfo {year} {1981})}\BibitemShut
  {NoStop}%
\bibitem [{\citenamefont {Stanton}\ and\ \citenamefont
  {Bartlett}(1993)}]{Stanton1993}%
  \BibitemOpen
  \bibfield  {author} {\bibinfo {author} {\bibfnamefont {J.~F.}\ \bibnamefont
  {Stanton}}\ and\ \bibinfo {author} {\bibfnamefont {R.~J.}\ \bibnamefont
  {Bartlett}},\ }\bibfield  {title} {\enquote {\bibinfo {title} {The equation
  of motion coupled-cluster method. a systematic biorthogonal approach to
  molecular excitation energies, transition probabilities, and excited state
  properties},}\ }\href {https://doi.org/10.1063/1.464746} {\bibfield
  {journal} {\bibinfo  {journal} {The Journal of Chemical Physics}\ }\textbf
  {\bibinfo {volume} {98}},\ \bibinfo {pages} {7029--7039} (\bibinfo {year}
  {1993})}\BibitemShut {NoStop}%
\bibitem [{\citenamefont {Krylov}(2008)}]{Krylov2008}%
  \BibitemOpen
  \bibfield  {author} {\bibinfo {author} {\bibfnamefont {A.~I.}\ \bibnamefont
  {Krylov}},\ }\bibfield  {title} {\enquote {\bibinfo {title}
  {Equation-of-motion coupled-cluster methods for open-shell and electronically
  excited species: The hitchhiker’s guide to fock space},}\ }\href
  {https://doi.org/10.1146/annurev.physchem.59.032607.093602} {\bibfield
  {journal} {\bibinfo  {journal} {Annual Review of Physical Chemistry}\
  }\textbf {\bibinfo {volume} {59}},\ \bibinfo {pages} {433--462} (\bibinfo
  {year} {2008})}\BibitemShut {NoStop}%
\bibitem [{\citenamefont {Bartlett}(2011)}]{Bartlett2011}%
  \BibitemOpen
  \bibfield  {author} {\bibinfo {author} {\bibfnamefont {R.~J.}\ \bibnamefont
  {Bartlett}},\ }\bibfield  {title} {\enquote {\bibinfo {title}
  {Coupled‐cluster theory and its equation‐of‐motion extensions},}\
  }\href {https://doi.org/10.1002/wcms.76} {\bibfield  {journal} {\bibinfo
  {journal} {WIREs Computational Molecular Science}\ }\textbf {\bibinfo
  {volume} {2}},\ \bibinfo {pages} {126--138} (\bibinfo {year}
  {2011})}\BibitemShut {NoStop}%
\bibitem [{\citenamefont {Sneskov}\ and\ \citenamefont
  {Christiansen}(2011)}]{Sneskov2011}%
  \BibitemOpen
  \bibfield  {author} {\bibinfo {author} {\bibfnamefont {K.}~\bibnamefont
  {Sneskov}}\ and\ \bibinfo {author} {\bibfnamefont {O.}~\bibnamefont
  {Christiansen}},\ }\bibfield  {title} {\enquote {\bibinfo {title} {Excited
  state coupled cluster methods},}\ }\href {https://doi.org/10.1002/wcms.99}
  {\bibfield  {journal} {\bibinfo  {journal} {WIREs Computational Molecular
  Science}\ }\textbf {\bibinfo {volume} {2}},\ \bibinfo {pages} {566--584}
  (\bibinfo {year} {2011})}\BibitemShut {NoStop}%
\bibitem [{\citenamefont {Cederbaum}, \citenamefont {Domcke},\ and\
  \citenamefont {Schirmer}(1980)}]{Cederbaum1980}%
  \BibitemOpen
  \bibfield  {author} {\bibinfo {author} {\bibfnamefont {L.~S.}\ \bibnamefont
  {Cederbaum}}, \bibinfo {author} {\bibfnamefont {W.}~\bibnamefont {Domcke}},\
  and\ \bibinfo {author} {\bibfnamefont {J.}~\bibnamefont {Schirmer}},\
  }\bibfield  {title} {\enquote {\bibinfo {title} {Many-body theory of core
  holes},}\ }\href {https://doi.org/10.1103/physreva.22.206} {\bibfield
  {journal} {\bibinfo  {journal} {Physical Review A}\ }\textbf {\bibinfo
  {volume} {22}},\ \bibinfo {pages} {206--222} (\bibinfo {year}
  {1980})}\BibitemShut {NoStop}%
\bibitem [{\citenamefont {Coriani}\ and\ \citenamefont
  {Koch}(2015{\natexlab{b}})}]{Coriani2016}%
  \BibitemOpen
  \bibfield  {author} {\bibinfo {author} {\bibfnamefont {S.}~\bibnamefont
  {Coriani}}\ and\ \bibinfo {author} {\bibfnamefont {H.}~\bibnamefont {Koch}},\
  }\bibfield  {title} {\enquote {\bibinfo {title} {Communication: {X}-ray
  absorption spectra and core-ionization potentials within a core-valence
  separated coupled cluster framework},}\ }\href
  {https://doi.org/10.1063/1.4935712} {\bibfield  {journal} {\bibinfo
  {journal} {The Journal of Chemical Physics}\ }\textbf {\bibinfo {volume}
  {143}},\ \bibinfo {pages} {181103} (\bibinfo {year}
  {2015}{\natexlab{b}})}\BibitemShut {NoStop}%
\bibitem [{\citenamefont {Vidal}\ \emph {et~al.}(2019)\citenamefont {Vidal},
  \citenamefont {Feng}, \citenamefont {Epifanovsky}, \citenamefont {Krylov},\
  and\ \citenamefont {Coriani}}]{Vidal2019}%
  \BibitemOpen
  \bibfield  {author} {\bibinfo {author} {\bibfnamefont {M.~L.}\ \bibnamefont
  {Vidal}}, \bibinfo {author} {\bibfnamefont {X.}~\bibnamefont {Feng}},
  \bibinfo {author} {\bibfnamefont {E.}~\bibnamefont {Epifanovsky}}, \bibinfo
  {author} {\bibfnamefont {A.~I.}\ \bibnamefont {Krylov}},\ and\ \bibinfo
  {author} {\bibfnamefont {S.}~\bibnamefont {Coriani}},\ }\bibfield  {title}
  {\enquote {\bibinfo {title} {New and efficient equation-of-motion
  coupled-cluster framework for core-excited and core-ionized states},}\ }\href
  {https://doi.org/10.1021/acs.jctc.9b00039} {\bibfield  {journal} {\bibinfo
  {journal} {Journal of Chemical Theory and Computation}\ }\textbf {\bibinfo
  {volume} {15}},\ \bibinfo {pages} {3117--3133} (\bibinfo {year}
  {2019})}\BibitemShut {NoStop}%
\bibitem [{\citenamefont {Tenorio}\ \emph {et~al.}(2019)\citenamefont
  {Tenorio}, \citenamefont {Moitra}, \citenamefont {Nascimento}, \citenamefont
  {Rocha},\ and\ \citenamefont {Coriani}}]{Tenorio2019}%
  \BibitemOpen
  \bibfield  {author} {\bibinfo {author} {\bibfnamefont {B.~N.~C.}\
  \bibnamefont {Tenorio}}, \bibinfo {author} {\bibfnamefont {T.}~\bibnamefont
  {Moitra}}, \bibinfo {author} {\bibfnamefont {M.~A.~C.}\ \bibnamefont
  {Nascimento}}, \bibinfo {author} {\bibfnamefont {A.~B.}\ \bibnamefont
  {Rocha}},\ and\ \bibinfo {author} {\bibfnamefont {S.}~\bibnamefont
  {Coriani}},\ }\bibfield  {title} {\enquote {\bibinfo {title} {Molecular
  inner-shell photoabsorption/photoionization cross sections at
  core-valence-separated coupled cluster level: Theory and examples},}\ }\href
  {https://doi.org/10.1063/1.5096777} {\bibfield  {journal} {\bibinfo
  {journal} {The Journal of Chemical Physics}\ }\textbf {\bibinfo {volume}
  {150}} (\bibinfo {year} {2019}),\ 10.1063/1.5096777}\BibitemShut {NoStop}%
\bibitem [{\citenamefont {Liu}\ \emph {et~al.}(2019)\citenamefont {Liu},
  \citenamefont {Matthews}, \citenamefont {Coriani},\ and\ \citenamefont
  {Cheng}}]{Liu2019}%
  \BibitemOpen
  \bibfield  {author} {\bibinfo {author} {\bibfnamefont {J.}~\bibnamefont
  {Liu}}, \bibinfo {author} {\bibfnamefont {D.}~\bibnamefont {Matthews}},
  \bibinfo {author} {\bibfnamefont {S.}~\bibnamefont {Coriani}},\ and\ \bibinfo
  {author} {\bibfnamefont {L.}~\bibnamefont {Cheng}},\ }\bibfield  {title}
  {\enquote {\bibinfo {title} {Benchmark calculations of k-edge ionization
  energies for first-row elements using scalar-relativistic
  core–valence-separated equation-of-motion coupled-cluster methods},}\
  }\href {https://doi.org/10.1021/acs.jctc.8b01160} {\bibfield  {journal}
  {\bibinfo  {journal} {Journal of Chemical Theory and Computation}\ }\textbf
  {\bibinfo {volume} {15}},\ \bibinfo {pages} {1642--1651} (\bibinfo {year}
  {2019})}\BibitemShut {NoStop}%
\bibitem [{\citenamefont {Myhre}, \citenamefont {Coriani},\ and\ \citenamefont
  {Koch}(2016)}]{Myhre2016}%
  \BibitemOpen
  \bibfield  {author} {\bibinfo {author} {\bibfnamefont {R.~H.}\ \bibnamefont
  {Myhre}}, \bibinfo {author} {\bibfnamefont {S.}~\bibnamefont {Coriani}},\
  and\ \bibinfo {author} {\bibfnamefont {H.}~\bibnamefont {Koch}},\ }\bibfield
  {title} {\enquote {\bibinfo {title} {Near-edge x-ray absorption fine
  structure within multilevel coupled cluster theory},}\ }\href
  {https://doi.org/10.1021/acs.jctc.6b00216} {\bibfield  {journal} {\bibinfo
  {journal} {Journal of Chemical Theory and Computation}\ }\textbf {\bibinfo
  {volume} {12}},\ \bibinfo {pages} {2633--2643} (\bibinfo {year}
  {2016})}\BibitemShut {NoStop}%
\bibitem [{\citenamefont {Frati}\ \emph {et~al.}(2019)\citenamefont {Frati},
  \citenamefont {de~Groot}, \citenamefont {Cerezo}, \citenamefont {Santoro},
  \citenamefont {Cheng}, \citenamefont {Faber},\ and\ \citenamefont
  {Coriani}}]{Frati2019}%
  \BibitemOpen
  \bibfield  {author} {\bibinfo {author} {\bibfnamefont {F.}~\bibnamefont
  {Frati}}, \bibinfo {author} {\bibfnamefont {F.}~\bibnamefont {de~Groot}},
  \bibinfo {author} {\bibfnamefont {J.}~\bibnamefont {Cerezo}}, \bibinfo
  {author} {\bibfnamefont {F.}~\bibnamefont {Santoro}}, \bibinfo {author}
  {\bibfnamefont {L.}~\bibnamefont {Cheng}}, \bibinfo {author} {\bibfnamefont
  {R.}~\bibnamefont {Faber}},\ and\ \bibinfo {author} {\bibfnamefont
  {S.}~\bibnamefont {Coriani}},\ }\bibfield  {title} {\enquote {\bibinfo
  {title} {Coupled cluster study of the x-ray absorption spectra of
  formaldehyde derivatives at the oxygen, carbon, and fluorine k-edges},}\
  }\href {https://doi.org/10.1063/1.5097650} {\bibfield  {journal} {\bibinfo
  {journal} {The Journal of Chemical Physics}\ }\textbf {\bibinfo {volume}
  {151}} (\bibinfo {year} {2019}),\ 10.1063/1.5097650}\BibitemShut {NoStop}%
\bibitem [{\citenamefont {Jensen}(2021)}]{Jensen2021}%
  \BibitemOpen
  \bibfield  {author} {\bibinfo {author} {\bibfnamefont {F.}~\bibnamefont
  {Jensen}},\ }\bibfield  {title} {\enquote {\bibinfo {title} {Computational
  chemistry: The exciting opportunities and the boring details},}\ }\href
  {https://doi.org/10.1002/ijch.202100027} {\bibfield  {journal} {\bibinfo
  {journal} {Israel Journal of Chemistry}\ }\textbf {\bibinfo {volume} {62}}
  (\bibinfo {year} {2021}),\ 10.1002/ijch.202100027}\BibitemShut {NoStop}%
\bibitem [{\citenamefont {Ireland}\ and\ \citenamefont
  {McKemmish}(2023)}]{Ireland2023}%
  \BibitemOpen
  \bibfield  {author} {\bibinfo {author} {\bibfnamefont {R.~T.}\ \bibnamefont
  {Ireland}}\ and\ \bibinfo {author} {\bibfnamefont {L.~K.}\ \bibnamefont
  {McKemmish}},\ }\bibfield  {title} {\enquote {\bibinfo {title} {On the
  specialization of gaussian basis sets for core-dependent properties},}\
  }\href {https://doi.org/10.1063/5.0159119} {\bibfield  {journal} {\bibinfo
  {journal} {The Journal of Chemical Physics}\ }\textbf {\bibinfo {volume}
  {159}} (\bibinfo {year} {2023}),\ 10.1063/5.0159119}\BibitemShut {NoStop}%
\bibitem [{\citenamefont {Watson}\ and\ \citenamefont
  {Bartlett}(2013)}]{Watson2013}%
  \BibitemOpen
  \bibfield  {author} {\bibinfo {author} {\bibfnamefont {T.~J.}\ \bibnamefont
  {Watson}}\ and\ \bibinfo {author} {\bibfnamefont {R.~J.}\ \bibnamefont
  {Bartlett}},\ }\bibfield  {title} {\enquote {\bibinfo {title} {Infinite order
  relaxation effects for core ionization energies with a variational coupled
  cluster ansatz},}\ }\href {https://doi.org/10.1016/j.cplett.2012.08.046}
  {\bibfield  {journal} {\bibinfo  {journal} {Chemical Physics Letters}\
  }\textbf {\bibinfo {volume} {555}},\ \bibinfo {pages} {235--238} (\bibinfo
  {year} {2013})}\BibitemShut {NoStop}%
\bibitem [{\citenamefont {Mijovilovich}\ \emph {et~al.}(2009)\citenamefont
  {Mijovilovich}, \citenamefont {Pettersson}, \citenamefont {Mangold},
  \citenamefont {Janousch}, \citenamefont {Susini}, \citenamefont {Salome},
  \citenamefont {de~Groot},\ and\ \citenamefont
  {Weckhuysen}}]{Mijovilovich2009}%
  \BibitemOpen
  \bibfield  {author} {\bibinfo {author} {\bibfnamefont {A.}~\bibnamefont
  {Mijovilovich}}, \bibinfo {author} {\bibfnamefont {L.~G.~M.}\ \bibnamefont
  {Pettersson}}, \bibinfo {author} {\bibfnamefont {S.}~\bibnamefont {Mangold}},
  \bibinfo {author} {\bibfnamefont {M.}~\bibnamefont {Janousch}}, \bibinfo
  {author} {\bibfnamefont {J.}~\bibnamefont {Susini}}, \bibinfo {author}
  {\bibfnamefont {M.}~\bibnamefont {Salome}}, \bibinfo {author} {\bibfnamefont
  {F.~M.~F.}\ \bibnamefont {de~Groot}},\ and\ \bibinfo {author} {\bibfnamefont
  {B.~M.}\ \bibnamefont {Weckhuysen}},\ }\bibfield  {title} {\enquote {\bibinfo
  {title} {The interpretation of sulfur k-edge xanes spectra: A case study on
  thiophenic and aliphatic sulfur compounds},}\ }\href
  {https://doi.org/10.1021/jp806823c} {\bibfield  {journal} {\bibinfo
  {journal} {The Journal of Physical Chemistry A}\ }\textbf {\bibinfo {volume}
  {113}},\ \bibinfo {pages} {2750--2756} (\bibinfo {year} {2009})}\BibitemShut
  {NoStop}%
\bibitem [{\citenamefont {Carbone}\ \emph {et~al.}()\citenamefont {Carbone},
  \citenamefont {Cheng}, \citenamefont {Myhre}, \citenamefont {Matthews},
  \citenamefont {Koch},\ and\ \citenamefont {Coriani}}]{Carbone2019}%
  \BibitemOpen
  \bibfield  {author} {\bibinfo {author} {\bibfnamefont {J.~P.}\ \bibnamefont
  {Carbone}}, \bibinfo {author} {\bibfnamefont {L.}~\bibnamefont {Cheng}},
  \bibinfo {author} {\bibfnamefont {R.~H.}\ \bibnamefont {Myhre}}, \bibinfo
  {author} {\bibfnamefont {D.}~\bibnamefont {Matthews}}, \bibinfo {author}
  {\bibfnamefont {H.}~\bibnamefont {Koch}},\ and\ \bibinfo {author}
  {\bibfnamefont {S.}~\bibnamefont {Coriani}},\ }\bibfield  {title} {\enquote
  {\bibinfo {title} {An analysis of the performance of coupled cluster methods
  for core excitations and core ionizations using standard basis sets},}\
  }\href {https://doi.org/10.1016/bs.aiq.2019.05.005} {\
  10.1016/bs.aiq.2019.05.005},\ \Eprint
  {https://arxiv.org/abs/http://arxiv.org/abs/1908.03635v1}
  {http://arxiv.org/abs/1908.03635v1} \BibitemShut {NoStop}%
\bibitem [{\citenamefont {Shirai}, \citenamefont {Yamamoto},\ and\
  \citenamefont {Hyodo}(2004)}]{Shirai2004}%
  \BibitemOpen
  \bibfield  {author} {\bibinfo {author} {\bibfnamefont {S.}~\bibnamefont
  {Shirai}}, \bibinfo {author} {\bibfnamefont {S.}~\bibnamefont {Yamamoto}},\
  and\ \bibinfo {author} {\bibfnamefont {S.-a.}\ \bibnamefont {Hyodo}},\
  }\bibfield  {title} {\enquote {\bibinfo {title} {Accurate calculation of
  core-electron binding energies: Multireference perturbation treatment},}\
  }\href {https://doi.org/10.1063/1.1799911} {\bibfield  {journal} {\bibinfo
  {journal} {The Journal of Chemical Physics}\ }\textbf {\bibinfo {volume}
  {121}},\ \bibinfo {pages} {7586--7594} (\bibinfo {year} {2004})}\BibitemShut
  {NoStop}%
\bibitem [{\citenamefont {Takahata}\ and\ \citenamefont
  {Chong}(2003)}]{Takahata2003}%
  \BibitemOpen
  \bibfield  {author} {\bibinfo {author} {\bibfnamefont {Y.}~\bibnamefont
  {Takahata}}\ and\ \bibinfo {author} {\bibfnamefont {D.~P.}\ \bibnamefont
  {Chong}},\ }\bibfield  {title} {\enquote {\bibinfo {title} {Dft calculation
  of core-electron binding energies},}\ }\href
  {https://doi.org/10.1016/j.elspec.2003.08.001} {\bibfield  {journal}
  {\bibinfo  {journal} {Journal of Electron Spectroscopy and Related
  Phenomena}\ }\textbf {\bibinfo {volume} {133}},\ \bibinfo {pages} {69--76}
  (\bibinfo {year} {2003})}\BibitemShut {NoStop}%
\bibitem [{\citenamefont {Wenzel}, \citenamefont {Wormit},\ and\ \citenamefont
  {Dreuw}(2014)}]{Wenzel2014}%
  \BibitemOpen
  \bibfield  {author} {\bibinfo {author} {\bibfnamefont {J.}~\bibnamefont
  {Wenzel}}, \bibinfo {author} {\bibfnamefont {M.}~\bibnamefont {Wormit}},\
  and\ \bibinfo {author} {\bibfnamefont {A.}~\bibnamefont {Dreuw}},\ }\bibfield
   {title} {\enquote {\bibinfo {title} {Calculating core‐level excitations
  and x‐ray absorption spectra of medium‐sized closed‐shell molecules
  with the algebraic‐diagrammatic construction scheme for the polarization
  propagator},}\ }\href {https://doi.org/10.1002/jcc.23703} {\bibfield
  {journal} {\bibinfo  {journal} {Journal of Computational Chemistry}\ }\textbf
  {\bibinfo {volume} {35}},\ \bibinfo {pages} {1900--1915} (\bibinfo {year}
  {2014})}\BibitemShut {NoStop}%
\bibitem [{\citenamefont {Kovač}\ \emph {et~al.}(2014)\citenamefont {Kovač},
  \citenamefont {Ljubić}, \citenamefont {Kivimäki}, \citenamefont {Coreno},\
  and\ \citenamefont {Novak}}]{Kovac2014}%
  \BibitemOpen
  \bibfield  {author} {\bibinfo {author} {\bibfnamefont {B.}~\bibnamefont
  {Kovač}}, \bibinfo {author} {\bibfnamefont {I.}~\bibnamefont {Ljubić}},
  \bibinfo {author} {\bibfnamefont {A.}~\bibnamefont {Kivimäki}}, \bibinfo
  {author} {\bibfnamefont {M.}~\bibnamefont {Coreno}},\ and\ \bibinfo {author}
  {\bibfnamefont {I.}~\bibnamefont {Novak}},\ }\bibfield  {title} {\enquote
  {\bibinfo {title} {Characterisation of the electronic structure of some
  stable nitroxyl radicals using variable energy photoelectron spectroscopy},}\
  }\href {https://doi.org/10.1039/c4cp00867g} {\bibfield  {journal} {\bibinfo
  {journal} {Phys. Chem. Chem. Phys.}\ }\textbf {\bibinfo {volume} {16}},\
  \bibinfo {pages} {10734--10742} (\bibinfo {year} {2014})}\BibitemShut
  {NoStop}%
\bibitem [{\citenamefont {Fransson}\ \emph {et~al.}(2016)\citenamefont
  {Fransson}, \citenamefont {Zhovtobriukh}, \citenamefont {Coriani},
  \citenamefont {Wikfeldt}, \citenamefont {Norman},\ and\ \citenamefont
  {Pettersson}}]{Fransson2016}%
  \BibitemOpen
  \bibfield  {author} {\bibinfo {author} {\bibfnamefont {T.}~\bibnamefont
  {Fransson}}, \bibinfo {author} {\bibfnamefont {I.}~\bibnamefont
  {Zhovtobriukh}}, \bibinfo {author} {\bibfnamefont {S.}~\bibnamefont
  {Coriani}}, \bibinfo {author} {\bibfnamefont {K.~T.}\ \bibnamefont
  {Wikfeldt}}, \bibinfo {author} {\bibfnamefont {P.}~\bibnamefont {Norman}},\
  and\ \bibinfo {author} {\bibfnamefont {L.~G.~M.}\ \bibnamefont
  {Pettersson}},\ }\bibfield  {title} {\enquote {\bibinfo {title} {Requirements
  of first-principles calculations of x-ray absorption spectra of liquid
  water},}\ }\href {https://doi.org/10.1039/c5cp03919c} {\bibfield  {journal}
  {\bibinfo  {journal} {Physical Chemistry Chemical Physics}\ }\textbf
  {\bibinfo {volume} {18}},\ \bibinfo {pages} {566--583} (\bibinfo {year}
  {2016})}\BibitemShut {NoStop}%
\bibitem [{\citenamefont {Tolbatov}\ and\ \citenamefont
  {Chipman}(2017)}]{Tolbatov2017}%
  \BibitemOpen
  \bibfield  {author} {\bibinfo {author} {\bibfnamefont {I.}~\bibnamefont
  {Tolbatov}}\ and\ \bibinfo {author} {\bibfnamefont {D.~M.}\ \bibnamefont
  {Chipman}},\ }\bibfield  {title} {\enquote {\bibinfo {title} {Benchmarking
  density functionals and gaussian basis sets for calculation of core-electron
  binding energies in amino acids},}\ }\href
  {https://doi.org/10.1007/s00214-017-2115-x} {\bibfield  {journal} {\bibinfo
  {journal} {Theoretical Chemistry Accounts}\ }\textbf {\bibinfo {volume}
  {136}} (\bibinfo {year} {2017}),\ 10.1007/s00214-017-2115-x}\BibitemShut
  {NoStop}%
\bibitem [{\citenamefont {Fouda}\ and\ \citenamefont
  {Besley}(2017)}]{Fouda2017}%
  \BibitemOpen
  \bibfield  {author} {\bibinfo {author} {\bibfnamefont {A.~E.~A.}\
  \bibnamefont {Fouda}}\ and\ \bibinfo {author} {\bibfnamefont {N.~A.}\
  \bibnamefont {Besley}},\ }\bibfield  {title} {\enquote {\bibinfo {title}
  {Assessment of basis sets for density functional theory-based calculations of
  core-electron spectroscopies},}\ }\href
  {https://doi.org/10.1007/s00214-017-2181-0} {\bibfield  {journal} {\bibinfo
  {journal} {Theoretical Chemistry Accounts}\ }\textbf {\bibinfo {volume}
  {137}} (\bibinfo {year} {2017}),\ 10.1007/s00214-017-2181-0}\BibitemShut
  {NoStop}%
\bibitem [{\citenamefont {Hanson-Heine}, \citenamefont {George},\ and\
  \citenamefont {Besley}(2018)}]{Hanson-Heine2018}%
  \BibitemOpen
  \bibfield  {author} {\bibinfo {author} {\bibfnamefont {M.~W.}\ \bibnamefont
  {Hanson-Heine}}, \bibinfo {author} {\bibfnamefont {M.~W.}\ \bibnamefont
  {George}},\ and\ \bibinfo {author} {\bibfnamefont {N.~A.}\ \bibnamefont
  {Besley}},\ }\bibfield  {title} {\enquote {\bibinfo {title} {Basis sets for
  the calculation of core-electron binding energies},}\ }\href
  {https://doi.org/10.1016/j.cplett.2018.03.066} {\bibfield  {journal}
  {\bibinfo  {journal} {Chemical Physics Letters}\ }\textbf {\bibinfo {volume}
  {699}},\ \bibinfo {pages} {279--285} (\bibinfo {year} {2018})}\BibitemShut
  {NoStop}%
\bibitem [{\citenamefont {Ambroise}\ and\ \citenamefont
  {Jensen}(2018)}]{Ambroise2018}%
  \BibitemOpen
  \bibfield  {author} {\bibinfo {author} {\bibfnamefont {M.~A.}\ \bibnamefont
  {Ambroise}}\ and\ \bibinfo {author} {\bibfnamefont {F.}~\bibnamefont
  {Jensen}},\ }\bibfield  {title} {\enquote {\bibinfo {title} {Probing basis
  set requirements for calculating core ionization and core excitation
  spectroscopy by the $\delta$ self-consistent-field approach},}\ }\href
  {https://doi.org/10.1021/acs.jctc.8b01071} {\bibfield  {journal} {\bibinfo
  {journal} {Journal of Chemical Theory and Computation}\ }\textbf {\bibinfo
  {volume} {15}},\ \bibinfo {pages} {325--337} (\bibinfo {year}
  {2018})}\BibitemShut {NoStop}%
\bibitem [{\citenamefont {Hait}\ and\ \citenamefont
  {Head-Gordon}(2020)}]{Hait2020}%
  \BibitemOpen
  \bibfield  {author} {\bibinfo {author} {\bibfnamefont {D.}~\bibnamefont
  {Hait}}\ and\ \bibinfo {author} {\bibfnamefont {M.}~\bibnamefont
  {Head-Gordon}},\ }\bibfield  {title} {\enquote {\bibinfo {title} {Highly
  accurate prediction of core spectra of molecules at density functional theory
  cost: Attaining sub-electronvolt error from a restricted open-shell
  kohn–sham approach},}\ }\href {https://doi.org/10.1021/acs.jpclett.9b03661}
  {\bibfield  {journal} {\bibinfo  {journal} {The Journal of Physical Chemistry
  Letters}\ }\textbf {\bibinfo {volume} {11}},\ \bibinfo {pages} {775--786}
  (\bibinfo {year} {2020})}\BibitemShut {NoStop}%
\bibitem [{\citenamefont {Cavigliasso}\ and\ \citenamefont
  {Chong}(1999)}]{Cavigliasso1999}%
  \BibitemOpen
  \bibfield  {author} {\bibinfo {author} {\bibfnamefont {G.}~\bibnamefont
  {Cavigliasso}}\ and\ \bibinfo {author} {\bibfnamefont {D.~P.}\ \bibnamefont
  {Chong}},\ }\bibfield  {title} {\enquote {\bibinfo {title} {Accurate
  density-functional calculation of core-electron binding energies by a
  total-energy difference approach},}\ }\href
  {https://doi.org/10.1063/1.480279} {\bibfield  {journal} {\bibinfo  {journal}
  {The Journal of Chemical Physics}\ }\textbf {\bibinfo {volume} {111}},\
  \bibinfo {pages} {9485--9492} (\bibinfo {year} {1999})}\BibitemShut {NoStop}%
\bibitem [{\citenamefont {Mejia-Rodriguez}\ \emph {et~al.}(2022)\citenamefont
  {Mejia-Rodriguez}, \citenamefont {Kunitsa}, \citenamefont {Aprà},\ and\
  \citenamefont {Govind}}]{Mejia-Rodriguez2022}%
  \BibitemOpen
  \bibfield  {author} {\bibinfo {author} {\bibfnamefont {D.}~\bibnamefont
  {Mejia-Rodriguez}}, \bibinfo {author} {\bibfnamefont {A.}~\bibnamefont
  {Kunitsa}}, \bibinfo {author} {\bibfnamefont {E.}~\bibnamefont {Aprà}},\
  and\ \bibinfo {author} {\bibfnamefont {N.}~\bibnamefont {Govind}},\
  }\bibfield  {title} {\enquote {\bibinfo {title} {Basis set selection for
  molecular core-level gw calculations},}\ }\href
  {https://doi.org/10.1021/acs.jctc.2c00247} {\bibfield  {journal} {\bibinfo
  {journal} {Journal of Chemical Theory and Computation}\ }\textbf {\bibinfo
  {volume} {18}},\ \bibinfo {pages} {4919--4926} (\bibinfo {year}
  {2022})}\BibitemShut {NoStop}%
\bibitem [{\citenamefont {Ambroise}, \citenamefont {Dreuw},\ and\ \citenamefont
  {Jensen}(2021)}]{Ambroise2021}%
  \BibitemOpen
  \bibfield  {author} {\bibinfo {author} {\bibfnamefont {M.~A.}\ \bibnamefont
  {Ambroise}}, \bibinfo {author} {\bibfnamefont {A.}~\bibnamefont {Dreuw}},\
  and\ \bibinfo {author} {\bibfnamefont {F.}~\bibnamefont {Jensen}},\
  }\bibfield  {title} {\enquote {\bibinfo {title} {Probing basis set
  requirements for calculating core ionization and core excitation spectra
  using correlated wave function methods},}\ }\href
  {https://doi.org/10.1021/acs.jctc.1c00042} {\bibfield  {journal} {\bibinfo
  {journal} {Journal of Chemical Theory and Computation}\ }\textbf {\bibinfo
  {volume} {17}},\ \bibinfo {pages} {2832--2842} (\bibinfo {year}
  {2021})}\BibitemShut {NoStop}%
\bibitem [{\citenamefont {Sarangi}\ \emph {et~al.}(2020)\citenamefont
  {Sarangi}, \citenamefont {Vidal}, \citenamefont {Coriani},\ and\
  \citenamefont {Krylov}}]{Sarangi2020}%
  \BibitemOpen
  \bibfield  {author} {\bibinfo {author} {\bibfnamefont {R.}~\bibnamefont
  {Sarangi}}, \bibinfo {author} {\bibfnamefont {M.~L.}\ \bibnamefont {Vidal}},
  \bibinfo {author} {\bibfnamefont {S.}~\bibnamefont {Coriani}},\ and\ \bibinfo
  {author} {\bibfnamefont {A.~I.}\ \bibnamefont {Krylov}},\ }\bibfield  {title}
  {\enquote {\bibinfo {title} {On the basis set selection for calculations of
  core-level states: different strategies to balance cost and accuracy},}\
  }\href {https://doi.org/10.1080/00268976.2020.1769872} {\bibfield  {journal}
  {\bibinfo  {journal} {Molecular Physics}\ }\textbf {\bibinfo {volume} {118}}
  (\bibinfo {year} {2020}),\ 10.1080/00268976.2020.1769872}\BibitemShut
  {NoStop}%
\bibitem [{\citenamefont {Dunning~Jr}(1989)}]{dunning1989gaussian}%
  \BibitemOpen
  \bibfield  {author} {\bibinfo {author} {\bibfnamefont {T.~H.}\ \bibnamefont
  {Dunning~Jr}},\ }\bibfield  {title} {\enquote {\bibinfo {title} {Gaussian
  basis sets for use in correlated molecular calculations. i. the atoms boron
  through neon and hydrogen},}\ }\href@noop {} {\bibfield  {journal} {\bibinfo
  {journal} {The Journal of chemical physics}\ }\textbf {\bibinfo {volume}
  {90}},\ \bibinfo {pages} {1007--1023} (\bibinfo {year} {1989})}\BibitemShut
  {NoStop}%
\bibitem [{\citenamefont {Kendall}, \citenamefont {Dunning},\ and\
  \citenamefont {Harrison}(1992)}]{kendall1992electron}%
  \BibitemOpen
  \bibfield  {author} {\bibinfo {author} {\bibfnamefont {R.~A.}\ \bibnamefont
  {Kendall}}, \bibinfo {author} {\bibfnamefont {T.~H.}\ \bibnamefont
  {Dunning}},\ and\ \bibinfo {author} {\bibfnamefont {R.~J.}\ \bibnamefont
  {Harrison}},\ }\bibfield  {title} {\enquote {\bibinfo {title} {Electron
  affinities of the first-row atoms revisited. systematic basis sets and wave
  functions},}\ }\href@noop {} {\bibfield  {journal} {\bibinfo  {journal} {The
  Journal of chemical physics}\ }\textbf {\bibinfo {volume} {96}},\ \bibinfo
  {pages} {6796--6806} (\bibinfo {year} {1992})}\BibitemShut {NoStop}%
\bibitem [{\citenamefont {Dunning~Jr}\ and\ \citenamefont
  {Woon}(1993)}]{dunning1993gaussian}%
  \BibitemOpen
  \bibfield  {author} {\bibinfo {author} {\bibfnamefont {T.}~\bibnamefont
  {Dunning~Jr}}\ and\ \bibinfo {author} {\bibfnamefont {D.}~\bibnamefont
  {Woon}},\ }\bibfield  {title} {\enquote {\bibinfo {title} {Gaussian basis
  sets for use in correlated molecular calculations. iii. the second row atoms,
  al--ar},}\ }\href@noop {} {\bibfield  {journal} {\bibinfo  {journal} {J.
  Chem. Phys}\ }\textbf {\bibinfo {volume} {98}},\ \bibinfo {pages}
  {1358--1371} (\bibinfo {year} {1993})}\BibitemShut {NoStop}%
\bibitem [{\citenamefont {Woon}\ and\ \citenamefont
  {Dunning~Jr}(1995)}]{woon1995gaussian}%
  \BibitemOpen
  \bibfield  {author} {\bibinfo {author} {\bibfnamefont {D.~E.}\ \bibnamefont
  {Woon}}\ and\ \bibinfo {author} {\bibfnamefont {T.~H.}\ \bibnamefont
  {Dunning~Jr}},\ }\bibfield  {title} {\enquote {\bibinfo {title} {Gaussian
  basis sets for use in correlated molecular calculations. v. core-valence
  basis sets for boron through neon},}\ }\href@noop {} {\bibfield  {journal}
  {\bibinfo  {journal} {The Journal of chemical physics}\ }\textbf {\bibinfo
  {volume} {103}},\ \bibinfo {pages} {4572--4585} (\bibinfo {year}
  {1995})}\BibitemShut {NoStop}%
\bibitem [{\citenamefont {Wilson}, \citenamefont {Van~Mourik},\ and\
  \citenamefont {Dunning~Jr}(1996)}]{wilson1996gaussian}%
  \BibitemOpen
  \bibfield  {author} {\bibinfo {author} {\bibfnamefont {A.~K.}\ \bibnamefont
  {Wilson}}, \bibinfo {author} {\bibfnamefont {T.}~\bibnamefont {Van~Mourik}},\
  and\ \bibinfo {author} {\bibfnamefont {T.~H.}\ \bibnamefont {Dunning~Jr}},\
  }\bibfield  {title} {\enquote {\bibinfo {title} {Gaussian basis sets for use
  in correlated molecular calculations. vi. sextuple zeta correlation
  consistent basis sets for boron through neon},}\ }\href@noop {} {\bibfield
  {journal} {\bibinfo  {journal} {Journal of Molecular Structure: THEOCHEM}\
  }\textbf {\bibinfo {volume} {388}},\ \bibinfo {pages} {339--349} (\bibinfo
  {year} {1996})}\BibitemShut {NoStop}%
\bibitem [{\citenamefont {Van~Mourik}, \citenamefont {Wilson},\ and\
  \citenamefont {Dunning~Jr}(1999)}]{van1999benchmark}%
  \BibitemOpen
  \bibfield  {author} {\bibinfo {author} {\bibfnamefont {T.}~\bibnamefont
  {Van~Mourik}}, \bibinfo {author} {\bibfnamefont {A.~K.}\ \bibnamefont
  {Wilson}},\ and\ \bibinfo {author} {\bibfnamefont {T.~H.}\ \bibnamefont
  {Dunning~Jr}},\ }\bibfield  {title} {\enquote {\bibinfo {title} {Benchmark
  calculations with correlated molecular wavefunctions. xiii. potential energy
  curves for he2, ne2 and ar2 using correlation consistent basis sets through
  augmented sextuple zeta},}\ }\href@noop {} {\bibfield  {journal} {\bibinfo
  {journal} {Molecular Physics}\ }\textbf {\bibinfo {volume} {96}},\ \bibinfo
  {pages} {529--547} (\bibinfo {year} {1999})}\BibitemShut {NoStop}%
\bibitem [{\citenamefont {Peterson}\ and\ \citenamefont
  {Dunning~Jr}(2002)}]{peterson2002accurate}%
  \BibitemOpen
  \bibfield  {author} {\bibinfo {author} {\bibfnamefont {K.~A.}\ \bibnamefont
  {Peterson}}\ and\ \bibinfo {author} {\bibfnamefont {T.~H.}\ \bibnamefont
  {Dunning~Jr}},\ }\bibfield  {title} {\enquote {\bibinfo {title} {Accurate
  correlation consistent basis sets for molecular core--valence correlation
  effects: The second row atoms al--ar, and the first row atoms b--ne
  revisited},}\ }\href@noop {} {\bibfield  {journal} {\bibinfo  {journal} {The
  Journal of chemical physics}\ }\textbf {\bibinfo {volume} {117}},\ \bibinfo
  {pages} {10548--10560} (\bibinfo {year} {2002})}\BibitemShut {NoStop}%
\bibitem [{\citenamefont {Matthews}\ \emph {et~al.}(2020)\citenamefont
  {Matthews}, \citenamefont {Cheng}, \citenamefont {Harding}, \citenamefont
  {Lipparini}, \citenamefont {Stopkowicz}, \citenamefont {Jagau}, \citenamefont
  {Szalay}, \citenamefont {Gauss},\ and\ \citenamefont
  {Stanton}}]{Matthews2020}%
  \BibitemOpen
  \bibfield  {author} {\bibinfo {author} {\bibfnamefont {D.~A.}\ \bibnamefont
  {Matthews}}, \bibinfo {author} {\bibfnamefont {L.}~\bibnamefont {Cheng}},
  \bibinfo {author} {\bibfnamefont {M.~E.}\ \bibnamefont {Harding}}, \bibinfo
  {author} {\bibfnamefont {F.}~\bibnamefont {Lipparini}}, \bibinfo {author}
  {\bibfnamefont {S.}~\bibnamefont {Stopkowicz}}, \bibinfo {author}
  {\bibfnamefont {T.-C.}\ \bibnamefont {Jagau}}, \bibinfo {author}
  {\bibfnamefont {P.~G.}\ \bibnamefont {Szalay}}, \bibinfo {author}
  {\bibfnamefont {J.}~\bibnamefont {Gauss}},\ and\ \bibinfo {author}
  {\bibfnamefont {J.~F.}\ \bibnamefont {Stanton}},\ }\bibfield  {title}
  {\enquote {\bibinfo {title} {Coupled-cluster techniques for computational
  chemistry: The cfour program package},}\ }\href
  {https://doi.org/10.1063/5.0004837} {\bibfield  {journal} {\bibinfo
  {journal} {The Journal of Chemical Physics}\ }\textbf {\bibinfo {volume}
  {152}} (\bibinfo {year} {2020}),\ 10.1063/5.0004837}\BibitemShut {NoStop}%
\end{thebibliography}%

\end{document}